# Multi-cell Hybrid Millimeter Wave Systems: Pilot Contamination and Interference Mitigation

Lou Zhao, *Student Member, IEEE*, Zhiqiang Wei, *Student Member, IEEE*, Derrick Wing Kwan Ng, *Senior Member, IEEE*, Jinhong Yuan, *Fellow, IEEE*, and Mark C. Reed *Senior Member, IEEE*

*Abstract*—**In this paper, we investigate the system performance of a multi-cell multi-user (MU) hybrid millimeter wave (mmWave) communications in a multiple-input multiple-output (MIMO) network. Due to the reuse of pilot symbols among different cells, the performance of channel estimation is expected to be degraded by pilot contamination, which is considered as a fundamental performance bottleneck of conventional multi-cell MU massive MIMO networks. To analyze the impact of pilot contamination to the system performance, we first derive the closed-form approximation of the normalized mean squared error (MSE) of the channel estimation algorithm proposed in [2] over Rician fading channels. Our analytical and simulation results show that the channel estimation error incurred by the impact of pilot contamination and noise vanishes asymptotically with an increasing number of antennas equipped at each radio frequency (RF) chain at the desired BS. Furthermore, by adopting zero-forcing (ZF) precoding in each cell for downlink transmission, we derive a tight closed-form approximation of the average achievable rate per user. Our results unveil that the intra-cell interference and inter-cell interference caused by pilot contamination over Rician fading channels can be mitigated effectively by simply increasing the number of antennas equipped at the desired BS.**

*Index Terms*—**Pilot contamination, millimeter wave, hybrid systems, inter-cell interference.**

## I. Introduction

Recently, the escalating demands for high data transmission, which is one of the key requirements of the fifth-generation (5G) wireless communication systems, have triggered and attracted tremendous interests from both academia and industry, e.g. [3]–[21]. To meet the ultra-high data rate requirement of emerging applications, millimeter wave (mmWave) massive multiple-input multiple-output (MIMO) systems have been proposed [22], [23]. Specifically, mmWave massive MIMO cellular systems provide a huge trunk of the unlicensed bandwidth of the order of gigahertz to achieve ultra-high data rate communication. In practice, to strike a balance between the data rate, hardware cost[1], system complexity, and power consumption, hybrid mmWave MIMO systems are proposed for practical implementation [8], [18], [22], [24]. Due to the hybrid MIMO structure, the required number of radio frequency (RF) chains equipped at a base station (BS) and users are much smaller than the number of antennas equipped at the BS and users. Thus, both the hardware cost and the energy consumption of hybrid mmWave massive MIMO systems can be reduced significantly compared to the conventional fully digital mmWave massive MIMO structure, e.g. [7], [8]. Furthermore, the severe propagation path loss of mmWave channels between the transceivers can be compensated by forming a highly directional information beam enabled by the massive numbers of antennas.

On the other hand, as required by 3GPP new radio (NR), the BSs of mmWave network systems is ultra-densely deployed with small cell radius for improving the network coverage and to meet the data rate requirement of the enhanced Mobile BroadBand (eMBB) applications (the peak data rate is 10 Gbps and the minimum is 100 Mbps[2]) [25], [26]. The ultra-dense mmWave network facilitates the reuse of the same piece of spectrum across a large geographic area[3] for achieving a high network spectral efficiency [27] and reducing the severe large-scale propagation path loss by shortening the distances between transceivers [12], [28]. In practice, the interference received at the desired user in ultra-dense mmWave network systems originates from two sources, as illustrated in Fig. 3 of [28]: interference among different BSs and interference within the desired cell. It is mentioned in [12] that, mmWave beams are highly directional, which completely changes the interference behavior as well as the sensitivity to beams misalignment. In particular, the interference adopts an on/off behavior where strong interference occurs intermittently [12]. With a shrinking cell radius, distances from neighbouring BSs to the desired user decrease, which may lead to severe inter-cell interference to the desired user. Recently, it is also mentioned in work [29] that the received interference significantly increases when

 L. Zhao, Z. Wei, D. W. K. Ng, and J. Yuan are with the School of Electrical Engineering and Telecommunications, University of New South Wales, Sydney, NSW 2052, Australia (Email: lou.zhao@unsw.edu.au; zhiqiang.wei@unsw.edu.au; w.k.ng@unsw.edu.au; j.yuan@unsw.edu.au). M. C. Reed is with the School of Engineering and Information Technology, University of New South Wales, Australian Defence Force Academy, Canberra, ACT 2610, Australia (Email: mark.reed@unsw.edu.au). This work was supported by the Australia Research Council Discovery Project under Grant DP160104566. D. W. K. Ng is supported by the Australian Research Council's Discovery Early Career Researcher Award funding scheme (DE170100137).

[1]The hardware cost is mainly contributed by RF, intermediate frequency (IF) and analog to digital converter/digital to analog converter (ADC/DAC) associated with each RF chain.

[2]Typical mmWave inter-site-distances (ISDs) would be 100-200 meters with experienced rates ranged from a few hundreds of Mbps to a few Gbps. In general, the peak data rate is not expected to be provided to every user in the cell.

[3]A reasonable frequency reuse factor can effectively strike a balance between cross-cell interference mitigation and bandwidth utilization. In this paper, we investigate the worst scenario in terms of multi-cell interference for mmWave networks that all cells reuse the same piece of frequency bands.

directional transmission are simultaneously adopted in both transceivers for the same amount of total emitted energy. Besides, due to the impact of imperfect channel state information (CSI) on the design of downlink precoder, the desired BS will cause severe intra-cell interference on the desired user. Thus, for the multi-small-cell mmWave downlink transmission with imperfect CSI and a large number of antennas, the desired user may suffer severe intra-cell and inter-cell interference. Hence, it is necessary to study the interference in multi-small-cell hybrid mmWave systems.

### A. Pilot Contamination in mmWave Systems

In mmWave systems, channel estimation is important for unlocking the potential of the system performance. Majority of contributions in the literature focus on the development of CSI feedback based channel estimation methods for hybrid mmWave systems, e.g. [20], [30]–[34]. In practice, the 5G NR specification includes a set of basic beam-related procedures for the above 6 GHz CSI acquisition. The beam management of 5G NR mmWave systems consists of four different operations, i.e., beam sweeping, beam measurement, beam determination, and beam reporting. Taking a single-antenna user in 5G NR stand-alone (SA) downlink scheme for example, the BS sequentially transmits synchronization signals (SS) and CSI reference signals (CSI-RSs) to the users by using a predefined codebook of directions. Then, the users search and track their optimal beams by measuring the collected CSI-RSs. At the end of CSI acquisition phase, beam reporting, the users feed back the determined beam information (beam quality and beam decision) and the random access channel (RACH) preamble to the BS.

However, these feedback-based channel estimation algorithms have some drawbacks. Most of channel estimation methods are based on the assumption of the sparsity of mmWave channels that the numbers of resolvable angles-of-arrival (AoA)/departure (AoD) paths are finite and limited. Thus, the CSI acquisition via feedback only leads to a small amount of signalling overhead compared to non-sparse CSI acquisition. However, the sparsity assumption for practical mmWave channels may not hold in some special scenarios[4] [18], [35]–[38]. Thus, the explicit CSI feedback from the users to the BS is required by the RACH-based approaches which may incur high complexity and extra signallings overhead. In particular, the required amount of feedback signalling overhead increases tremendously with the number of scattering components as well as scales with the number of antennas equipped at the BS, consuming a significant portion of system resources [30]. In addition, there will be a system rate performance degradation due to the limited amount of the feedback and the limited resolution of CSI quantization [30]. Therefore, a low computational complexity mmWave channel estimation algorithm, which does not require explicit CSI feedback, is necessary to unlock the potential of mmWave massive MIMO systems. Recently, an alternative algorithm, which performs the mmWave channel estimation in the uplink by exploiting the reciprocity of downlink in TDD systems and sound reference signal (SRS), has been investigated in works [2], [39]. The utilization of SRS, e.g. orthogonal pilot symbols, transmitted from the users to the desired BS for the estimation of equivalent channels is novel and different from previous works utilizing SS and CSI-RS [40]–[42]. The analytical and numerical results illustrated that mmWave multi-user (MU) channel estimation at the BS in uplink outperforms downlink channel estimation at the users for a fixed overhead, even if the required signalling overhead feedback is not taken into account [39]. The results hold for both fully digital and hybrid architectures at the receiver and transmitter sides and have been verified via the recent field test [43]. Besides, the algorithm mentioned above can work well in sparse and non-sparse environments[5] [2].

For the algorithm proposed in [2], the designed analog beamformers allow signal transmission and reception along the strongest AoA directions, which reduce the interference outside the strongest AoA directions and utilize the transmission power more efficiently. It can strike a balance between the system robustness [41], the impact of multi-paths on signal processing[6], and the receive desired energy degradation of the received signal [2]. It was shown that the algorithm proposed in [2] could achieve a considerable achievable rate of the optimal fully digital systems and possess robustness against beam misalignments as well as the hardware imperfection. Besides, the third step of the algorithm proposed in [2] can be applied to the existing beam sweeping protocols in 5G NR for further evolution. Yet, paper [2] only considered a simple single-cell scenario. Note that, the introduction of frequency reuse and small-cell incur strong inter-cell interference during the uplink channel estimation phase and the downlink data transmission phase [12], [17], [22], [29].

### B. Related Prior Works

Since the resources of orthogonal pilots are limited, they are reused among different cells for multi-cell channel estimation. In this case, the received pilot symbols from the users in the desired cell for channel estimation are affected by the reused pilot symbols from the users in neighboring cells, which is known as pilot contamination [17], [46]. As a result, the downlink transmission based on the CSI obtained via contaminated pilots causes severe intra-cell and inter-cell interference in the desired cell. In fact, pilot contamination is considered as a fundamental performance bottleneck of the conventional multi-cell multiuser (MU) massive MIMO systems, since the resulting channel estimation errors do not vanish even if the number of antennas is sufficiently large, cf. [4], [17], [46]–[48]. Recently, various algorithms [46], [49]–[54] have been proposed to alleviate the impact of pilot contamination, e.g. data-aided iterative channel estimation algorithms, pilot design algorithms, multi-cell minimum mean squared error (MMSE) based precoding algorithms, and so forth. However, some algorithms, e.g. multi-cell MMSE algorithm, are mostly based on the assumption that the desired BS can have perfect knowledge of covariance matrices of pilot-sharing users in neighbouring cells, which is overly optimistic. Besides, the condition that the desired BS has the perfect knowledge of covariance matrices is a necessary but not sufficient condition for pilot contamination mitigation, cf. [52]–[54]. The algorithm proposed in [52] requires that covariance matrices of pilot-sharing users in neighbouring cells are orthogonal, which is unlikely in practice [54]. Also, the requirement of [53] for completely eliminating pilot contamination is that the number of antennas equipped at the BS and the size of a coherence time block jointly go to infinity.

[4]For certain practical mmWave channels, there are not enough measurement data for justifying that mmWave channels in some considered scenarios are non-sparse yet. However, recent field test results, as well as ray-tracing simulation results, have shown that the number of scattering clusters increases significantly and reflections from street signs, lamp posts, parked vehicles, passing people, etc., could reach a receiver from all possible directions in UMi scenarios in the city center [18], [35], [36].

[5]The algorithm proposed in [2] can exploit orthogonal pilot symbols transmission from users to the BS via the strongest received AoA paths in the case of line-of-sight (LOS) and non-line-of-sight (NLOS) environments. The NLOS scenario is considered as the most common user-case of mmWave systems, cf., the 3GPP channel model as well as various discussions on the development in the channel modeling literature e.g. [18], [38], [44] .

[6]The after-analog-beamforming root-mean-square (RMS) delay spread of power delay profile (PDP) of equivalent channels is significantly smaller than that of omni-directional antenna estimated exact channels [38], [44], [45].

In the literature, most of existing multi-cell massive MU-MIMO works for pilot contamination [12], [17], [46], [47] assumed that cross-cell channels from pilot-sharing users in neighbouring cells to the desired BS are Rayleigh fading channels with zero means. However, as discussed in Section IV of [55], it is probably not accurate for modeling the inter-cell interference as a Gaussian random variable with a small-cell setting. In fact, recent field measurements have confirmed that the strongest AoA components always exist in the inter-cell mmWave channels in small-cell systems [18], [35], [36], [38]. Besides, the mean values of cross-cell channels are not zero and different from each other. In other words, the distribution of cross-cell mmWave channels is different from that of the sub-6 GHz channels. Thus, the results obtained in works mentioned above for pilot contamination mitigation and performance analysis, e.g. [12], [17], [22], [46], [54], cannot be applied directly. Furthermore, a thorough study on the impact of pilot contamination in such a practical network system has not been reported yet.

### C. Contributions

Motivated by the aforementioned discussions, we consider a multi-small-cell MU hybrid mmWave system. In particular, we apply the non-feedback TDD-based mmWave channel estimation algorithm proposed in [2] to the considered multi-cell scenario and study the corresponding performance of channel estimation under the impact of pilot contamination. In addition, we analyze the downlink achievable rate performance of the hybrid mmWave system by taking into account the CSI errors caused by pilot contamination and inter-cell interferences originated from the BSs in neighboring cells. Note that, for pilot contamination mitigation, this paper does not require any information of covariance matrices of pilot-sharing users in neighbouring cells as required by the multi-cell MMSE-based precoding algorithm proposed in [54]. In addition, the algorithm adopted in this paper can work well while mean values of cross-cell channels from pilot-sharing users in neighbouring cells to the desired BS are non-zero. Our main contributions are summarized as follows:

- We apply the three-step mmWave channel estimation algorithm proposed in [2] to a multi-cell scenario. Then, we study the impact of pilot contamination on the uplink mmWave channel estimation due to the reuse of orthogonal pilot symbols among different cells. Our results reveal that in the phase of channel estimation, the receive analog beamforming matrix adopted at the desired BS forms a spatial filter which blocks the signal reception of the undesired pilot symbols from neighboring cells. In particular, with an increasing number of antennas equipped at each RF chain, the mainlobe beamwidth of the spatial filter, which aligns to the strongest AoA direction, becomes narrower and the amplitude of sidelobes becomes lower. Thus, the impact of pilot contamination caused by the users outside strongest AoA directions can be mitigated. We mathematically prove that the normalized mean squared error (MSE) performance of the channel estimation algorithm proposed in [2] improves proportionally with the increasing number of antennas equipped at each RF chain, which is different from previous results in [17], [46].
- We adopt zero-forcing (ZF) precoding for the downlink transmission based on the estimated CSI. Taking into account the impact of pilot contamination and the inter-cell interference from the neighboring BSs, we analyze and derive the closed-form approximation of the average achievable rate performance in the large number of antennas regimes. In addition, we show that the achievable rate scales with $M$ in the multi-cell scenario, where $M$ is the number of antennas equipped at each RF chain. We derive a scaling law of the achievable rate per users, $\log_2 M$. In contrast, the average achievable rate of fully digital systems adopting the conventional least squares (LS) channel estimation algorithm suffers from the severe impact of pilot contamination. All the derived analytical results are verified via simulations.

Notation: $\mathbb{E}_{\mathrm{h}}(\cdot)$ denotes statistical expectation operation with respect to random variable $h$, $\mathbb{C}^{M\times N}$ denotes the space of all $M \times N$ matrices with complex entries; $(\cdot)^{-1}$ denotes inverse operation; $(\cdot)^H$ denotes Hermitian transpose; $(\cdot)^*$ denotes complex conjugate; $(\cdot)^T$ denotes transpose; $|\cdot|$ denote the absolute value of a complex scalar; $\mathrm{tr}(\cdot)$ denotes trace operation; $\mathrm{sinc}(x)$ denotes a sinc function with input $x$, i.e., $\frac{\sin(x)}{x}$. The distribution of a circularly symmetric complex Gaussian (CSCG) random vector with a mean vector $\mathbf{x}$ and a covariance matrix $\sigma^2\mathbf{I}$ is denoted by $\mathcal{CN}(\mathbf{x}, \sigma^2\mathbf{I})$, and $\sim$ means "distributed as". $\mathbf{I}_{\mathrm{P}}$ is an $P \times P$ identity matrix.

## II. System Model

In this paper, a multi-cell MU hybrid subarray mmWave system is considered. The system consists of $L$ neighboring cells and there are one BS and $N$ users in each cell, cf. Fig. 1. The BS in each cell is equipped with $N_{\mathrm{RF}}$ RF chains serving the $N$ users simultaneously. We assume that each RF chain equipped at the BS can access to a uniform linear array (ULA) with $M$ antennas by using $M$ phase shifters. Besides, each user is equipped with one RF chain and a $P$-antenna array. In addition, we focus on $M \geqslant N_{\mathrm{RF}}$, which exploits a large antenna array gain with limited number of RF chains. Work [10] proved that when the number of RF chains equipped at the BS is more than twice of the number of users, the hybrid systems can realize the rate performance of the fully digital systems. In general, for the case of $N_{\mathrm{RF}} = N$, the hybrid system can provide enough degrees of freedom to serve $N$ users simultaneously. If the BS can equip with more RF chains, more degrees of freedom can be exploited for the design of analog beamformers and digital precoders. In addition, the extra RF chains adopted by the BS can be used for user scheduling design and multipath diversity exploitation. Thus, for the case that $N_{\mathrm{RF}} > N$, hybrid systems can achieve higher rate performance than that of $N_{\mathrm{RF}} = N$. In addition, authors of [56]–[58] employed RF chains with sets of digitally controlled phase paired phase-shifters and switches to achieve the same rate performance as that of the fully digital system. Furthermore, with the support of an advanced hybrid architecture [56]–[58], hybrid systems can achieve higher rate performance than that of adopting a simple ULA antenna array case. However, how to exploit extra RF chains as well as advanced hybrid architectures to further improve the rate performance of hybrid systems is beyond the scope of this paper. We may consider the case of $N_{\mathrm{RF}} > N$ and the case of adopting advanced hybrid architectures for channel estimation and downlink transmission precoding in our future works. To simplify the analysis in the following sections, without loss of generality, we set $N_{\mathrm{RF}} = N$ and each cell has the same number of RF chains equipped at the BS.

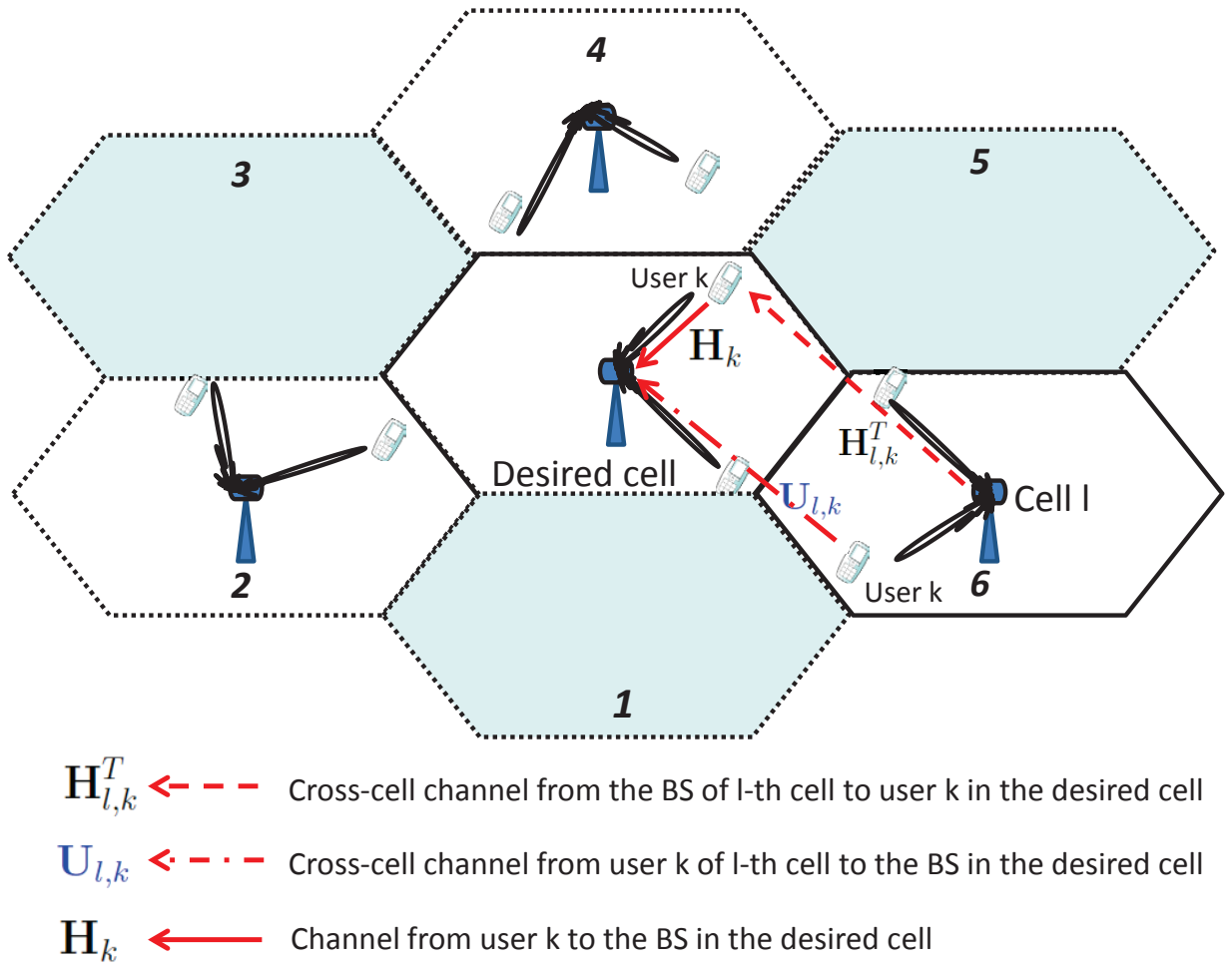

Fig. 1. A multi-cell MU mmWave cellular system with $L = 6$ neighboring cells.

According to the widely adopted setting for multi-cell TDD in uplink channel estimation and downlink data transmission, we assume that the users and the BSs in all cells are fully synchronized in time [17], [46]. We denote $\mathbf{H}_k \in \mathbb{C}^{M\times P}$ as the uplink channel matrix between the desired BS and user $k$ in the desired cell. Besides, $\mathbf{H}_k$ is a narrowband slow time-varying block fading channel. Recent field tests show that both strong LOS components and non-negligible scattering components may exist in mmWave propagation channels [18], [36], especially in the urban areas. In this paper, without loss of generality, $\mathbf{H}_k$, consists of a strongest AoA component $\mathbf{H}_{\mathrm{SAoA},k} \in \mathbb{C}^{M\times P}$ and $N_{\mathrm{cl}}$ scattering clusters/paths $\mathbf{H}_{\mathrm{S},k,i} \in \mathbb{C}^{M\times P}$, $i \in \{1,\cdots,N_{\mathrm{cl}}\}$ [38], [40], [41], [59], [60], which can be expressed as

$$\mathbf{H}_k = \frac{\sqrt{\varpi_k}\left[\alpha_{k,0}\mathbf{H}_{\mathrm{SAoA},k}+\sqrt{\frac{1}{N_{\mathrm{cl}}}}\sum\limits_{i=1}^{N_{\mathrm{cl}}}\alpha_{k,i}\mathbf{H}_{\mathrm{S},k,i}\right]}{\sqrt{|\alpha_{k,0}|^2+\frac{1}{N_{\mathrm{cl}}}\sum\limits_{i=1}^{N_{\mathrm{cl}}}|\alpha_{k,i}|^2}}, \tag{1}$$

where $\varpi_k$ accounts for the corresponding large-scale path loss, $\alpha_{k,0}$ is the complex gain corresponding to the strongest AoA component and $\alpha_{k,i} \sim \mathcal{CN}(0,1)$, $i \in \{1,\cdots,N_{\mathrm{cl}}\}$ represents the complex gain corresponding to the $i$-th cluster. We can also assume that $\{\alpha_{k,i}\}$ are in non-increasing order, i.e., $|\alpha_{k,0}| \geq |\alpha_{k,1}| \geq \cdots \geq |\alpha_{k,N_{\mathrm{cl}}}|$. We note here, the strongest AoAs can be either from the LOS components or the NLOS components. Then, the strongest AoA component of user $k$ in the desired cell, $\mathbf{H}_{\mathrm{SAoA},k}$, can be expressed as [61]

$$\mathbf{H}_{\mathrm{SAoA},k} = \mathbf{h}_{\mathrm{SAoA},k}^{\mathrm{BS}}\mathbf{h}_{\mathrm{SAoA},k}^{H}, \tag{2}$$

where $\mathbf{h}_{\mathrm{SAoA},k}^{\mathrm{BS}} \in \mathbb{C}^{M\times 1}$ and $\mathbf{h}_{\mathrm{SAoA},k} \in \mathbb{C}^{P\times 1}$ are the antenna array response vectors of the BS and user $k$, respectively. In particular, $\mathbf{h}_{\mathrm{SAoA},k}^{\mathrm{BS}}$ and $\mathbf{h}_{\mathrm{SAoA},k}$ can be expressed as [61]

$$\mathbf{h}_{\mathrm{SAoA},k}^{\mathrm{BS}} = \left[\begin{array}{c} 1,\ldots,e^{-j2\pi(M-1)\frac{d}{\lambda}\cos(\theta_k)} \end{array}\right]^T \text{ and} \tag{3}$$

$$\mathbf{h}_{\mathrm{SAoA},k} = \left[\begin{array}{c} 1,\ldots,e^{-j2\pi(P-1)\frac{d}{\lambda}\cos(\phi_k)} \end{array}\right]^T, \tag{4}$$

respectively, where $d$ is the distance between the neighboring antennas at the BS and users and $\lambda$ is the wavelength of the carrier frequency. Variables $\theta_k \in [0,\ \pi]$ and $\phi_k \in [0,\ \pi]$ are the angle of incidence of the strongest path at antenna arrays of the desired BS and user $k$, respectively. As commonly adopted in the literature [61], we set $d = \frac{\lambda}{2}$ for convenience. Similarly, the scattering components of user $k$ in the desired cell, $\mathbf{H}_{\mathrm{S},k}$, can be expressed as

$$\mathbf{H}_{\mathrm{S},k} = \sqrt{\frac{1}{N_{\mathrm{cl}}}}\sum_{i=1}^{N_{\mathrm{cl}}}\alpha_{k,i}\mathbf{H}_{\mathrm{S},k,i} = \sqrt{\frac{1}{N_{\mathrm{cl}}}}\sum_{i=1}^{N_{\mathrm{cl}}}\alpha_{k,i}\mathbf{h}_{k,i}^{\mathrm{BS}}\mathbf{h}_{k,i}^{H}, \tag{5}$$

where $\mathbf{h}_{k,i}^{\mathrm{BS}} \in \mathbb{C}^{M\times 1}$ and $\mathbf{h}_{k,i} \in \mathbb{C}^{P\times 1}$ are the antenna array response vectors of the BS and user $k$ associated to the $i$-th propagation path, respectively. By introducing the power ratio of the strongest cluster power over the other scattered paths of user $k$[7], $\varsigma_k = \frac{|\alpha_{k,0}|^2}{\frac{1}{N_{\mathrm{cl}}}\sum\limits_{i=1}^{N_{\mathrm{cl}}}|\alpha_{k,i}|^2}$, we can rewrite Equation (1) as

$$\mathbf{H}_k = \underbrace{\sqrt{\frac{\varpi_k\varsigma_k}{\varsigma_k+1}}\mathbf{h}_{\mathrm{SAoA},k}^{\mathrm{BS}}\mathbf{h}_{\mathrm{SAoA},k}^{H}}_{\text{Strongest AoA}}$$

[7]According to field test results summarized in Table I of [38] in the case of LOS and NLOS environments, the power ratio of the strongest cluster power over the other scattered paths' power, $\varsigma_k$, is larger than 1.

$$+\underbrace{\sqrt{\frac{\varpi_k}{\varsigma_k+1}}\sqrt{\frac{1}{N_{\mathrm{cl}}}}\sum_{i=1}^{N_{\mathrm{cl}}}\alpha_{k,i}\mathbf{h}_{k,i}^{\mathrm{BS}}\mathbf{h}_{k,i}^{H}}_{\text{Scattering components}}. \tag{6}$$

With the increasing number of clusters, the path attenuation coefficients and the AoAs between the users and the BS behave randomly [36]. Let $\mathbf{U}_{l,k} \in \mathbb{C}^{M\times P}$ be the inter-cell mmWave uplink channel between user $k$ in the $l$-th neighboring cell and the desired BS, $l \in \{1,\ldots,L\}$, cf. Fig. 1. Let $\mathbf{H}_{l,k}^{T} \in \mathbb{C}^{P\times M}$ be the inter-cell downlink mmWave channel between the BS of the $l$-th neighboring cell and user $k$ in the desired cell, cf. Fig. 1. Since the inter-cell distance is shortened in small-cell systems, the inter-cell channels usually contain the strongest AoA components. Thus, the inter-cell uplink and downlink mmWave channels can be expressed as

$$\mathbf{U}_{l,k} = \underbrace{\sqrt{\frac{\overline{\varpi}_{l,k}\varsigma_{l,k}}{\varsigma_{l,k}+1}}\mathbf{U}_{\mathrm{SAoA},l,k}}_{\text{Strongest AoA}} + \underbrace{\sqrt{\frac{\overline{\varpi}_{l,k}}{\varsigma_{l,k}+1}}\mathbf{U}_{\mathrm{S},l,k}}_{\text{Scattering component}} \quad \text{and} \tag{7}$$

$$\mathbf{H}_{l,k}^{T} = \underbrace{\sqrt{\frac{\widetilde{\varpi}_{l,k}\widetilde{\varsigma}_{l,k}}{\widetilde{\varsigma}_{l,k}+1}}\mathbf{H}_{\mathrm{SAoA},l,k}^{T}}_{\text{Strongest AoA}} + \underbrace{\sqrt{\frac{\widetilde{\varpi}_{l,k}}{\widetilde{\varsigma}_{l,k}+1}}\mathbf{H}_{\mathrm{S},l,k}^{T}}_{\text{Scattering component}}, \tag{8}$$

respectively, where $\overline{\varpi}_{l,k}$ and $\widetilde{\varpi}_{l,k}$ are the corresponding large-scale path loss coefficients. We note that $\varsigma_{l,k}$ and $\widetilde{\varsigma}_{l,k}$ are the power ratios of the strongest cluster power over the other scattered paths' power for the uplink and downlink inter-cell channels, respectively. The strongest AoA components $\mathbf{U}_{\mathrm{SAoA},l,k} = \mathbf{u}_{\mathrm{SAoA},l,k}^{\mathrm{BS}}\mathbf{u}_{\mathrm{SAoA},l,k}^{H}$, and $\mathbf{H}_{\mathrm{SAoA},l,k}^{T} = \mathbf{h}_{\mathrm{SAoA},l,k}^{*}\left(\mathbf{h}_{\mathrm{SAoA},l,k}^{\mathrm{BS}}\right)^{T}$ follow similar assumptions as in Equations $(2)-(4)$. Besides, the scattering components $\mathbf{U}_{\mathrm{S},l,k}$ and $\mathbf{H}_{\mathrm{S},l,k}^{T}$ follow similar assumption as in Equation (5). All these mentioned inter-cell propagation path loss coefficients are related to the propagation distance and modeled as in [18], [35], [36]. According to recent field measurements, e.g. [18], [35], [36], [62], [63], the typical values of the power ratios of the strongest cluster power over the other scattered paths' power $\varsigma_{l,k}$ and $\widetilde{\varsigma}_{l,k}$ for $\mathbf{U}_{l,k}$ and $\mathbf{H}_{l,k}^{T}$ are in $[0,\ 5]$, respectively.

## III. Multi-cell Uplink Channel Estimation Performance Analysis

In this section, we adopt the algorithm proposed in [2] for the estimation of an equivalent mmWave channel which comprise the physical mmWave channels and analog beamforming matrices adopted at the desired transceivers. The proposed algorithm is suitable for both the conventional fully digital systems and the emerging hybrid systems with fully access and subarray implementation structures. For the sake of presentation, we provide a summary of the algorithm proposed in [2] in the following. Specifically, each RF chain can access to the $M$ antennas via a phase shifter network.

There are three steps in the algorithm proposed in [2]. In the first and second steps, the strongest AoAs between the desired BS and the users are estimated. Then, the desired BS and the users design their analog beamforming matrices by aligning the beamforming direction to the estimated strongest AoAs, which is similar to [64]. In the third step, orthogonal pilot symbols are transmitted from the users to the BS by using the pre-designed analog beamforming matrices. Then, by exploiting the channel reciprocity, the equivalent downlink channel can be estimated and adopted at the BS as the input for the digital baseband precoder.

Note that, orthogonal pilot symbols are used for the estimation of the equivalent channels in the third step. For the multi-cell scenario, pilot symbols are reused among different cells which results in pilot contamination and cause a severe impact on the equivalent channel estimation performance. We note that only single-cell scenario was considered in [2] and it is unclear if the channel estimation algorithm provides robustness against pilot contamination. Furthermore, the performance analysis studied in [2] does not take into account the impact of potential out-of-cell interference on the performance of channel estimation. In the following sections, we investigate the impact of pilot contamination on the mmWave channel estimation performance for small-cell scenarios.

### A. Channel Estimation and Pilot Contamination Analysis

Basically, step one and step two of the proposed algorithm in [2] provide the analog beamforming matrices at the desired BS and the desired users to facilitate the estimation of equivalent channel. In particular, the analog beamforming matrices pair the desired BS and the users and align the directions of data stream transmission. Due to the inter-cell large-scale propagation path loss, the impact of multi-cell interference on the design of the analog beamforming matrices is usually negligible[8]. Besides, simulation results show that the pilot contamination does not affect strongest AoAs estimation result[9]. Therefore, to facilitate the performance analysis of the multi-cell equivalent channel estimation, we assume that strongest AoAs among the users and the BS are perfectly estimated and the desired signals always fall in the mainlobe.

Based on the strongest AoAs, which are perfectly estimated at the users and the BS, the analog receive beamforming vector of user $k$ adopted at the desired BS is given by

$$\widehat{\boldsymbol{\nu}}_k^{T} \in \mathbb{C}^{1\times M} = \frac{1}{\sqrt{M}}\left[\ 1,\ldots,e^{j2\pi(M-1)\frac{d}{\lambda}\cos(\theta_k)}\ \right] \tag{9}$$

and the analog transmit beamformer of user $i$ in the desired cell is given by

$$\widehat{\boldsymbol{\omega}}_i^{*} \in \mathbb{C}^{P\times 1} = \frac{1}{\sqrt{P}}\left[\ 1,\ldots,e^{j2\pi(P-1)\frac{d}{\lambda}\cos(\phi_i)}\ \right]^{H}. \tag{10}$$

In addition, we denote the analog beamforming matrix at the desired BS as

$$\mathbf{F}_{\mathrm{RF}} \in \mathbb{C}^{M\times N} = \left[\ \widehat{\boldsymbol{\nu}}_1,\ldots,\widehat{\boldsymbol{\nu}}_N\ \right]. \tag{11}$$

[8] For strongest AoAs estimation at the desired BS in a multi-cell scenario, the received power of the reused pilot symbols transmitted from the users in neighboring cells is smaller than that of the desired pilot symbols transmitted from the users in the desired cell. In addition, strongest AoAs estimation may not rely on pilot symbols [23].

[9] The simulation result of the strongest AoAs estimation is omitted here due to the space limitation. However, the impact of pilot contamination on the strongest AoAs estimation will be captured in the final simulation.

Let $\mathbf{\Phi}_k \in \mathbb{C}^{N\times 1}$ denote the pilot symbols of user $k$ in the desired cell. The pilot symbols for all the $N$ users in the desired cell form a matrix, $\mathbf{\Psi} \in \mathbb{C}^{N\times N}$, where $\mathbf{\Phi}_k$ is a column vector of matrix $\mathbf{\Psi}$ given by $\mathbf{\Psi} = \sqrt{E_{\mathrm{P}}}\left[\begin{array}{c} \mathbf{\Phi}_1,\ldots,\mathbf{\Phi}_N \end{array}\right]$, $\mathbf{\Phi}_i^H\mathbf{\Phi}_j = 0,\ i\neq j,\ i,\ j\in\{1,\ldots,N\}$, where $E_{\mathrm{P}}$ represents the transmitted pilot symbol energy.

In the equivalent channel estimation, user $k$ transmits the pilot symbols $\mathbf{\Phi}_k$ via transmit beamformer $\widehat{\boldsymbol{\omega}}_i^*$ and the desired BS receives the pilot symbols utilizing the analog beamforming matrix $\mathbf{F}_{\mathrm{RF}}$. However, the reuse of pilot symbols in neighboring cells affects the performance of equivalent channel estimation. The received signal of the $k$-th RF chain at the desired BS in the uplink is given by

$$\begin{aligned}\widehat{\mathbf{s}}_k^T =& \widehat{\boldsymbol{\nu}}_k^T\sum_{i=1}^{N}\mathbf{H}_i\widehat{\boldsymbol{\omega}}_i^*\sqrt{E_{\mathrm{P}}}\mathbf{\Phi}_i^T \\ &+\underbrace{\widehat{\boldsymbol{\nu}}_k^T\sum_{l=1}^{L}\sum_{i=1}^{N}\left(\mathbf{U}_{l,i}\widehat{\boldsymbol{\omega}}_{l,i}^*\sqrt{E_{\mathrm{P}}}\mathbf{\Phi}_i^T\right)}_{\text{Pilot contamination}}+\widehat{\boldsymbol{\nu}}_k^T\mathbf{Z},\end{aligned} \tag{12}$$

where $\widehat{\boldsymbol{\omega}}_{l,i} \in \mathbb{C}^{P\times 1} = \frac{1}{\sqrt{P}}\left[\begin{array}{c} 1,\ldots,e^{j2\pi(P-1)\frac{d}{\lambda}\cos(\phi_{l,k})} \end{array}\right]^T$ is the analog beamforming vector of user $i$ in the $l$-th cell, the entries of noise matrix, $\mathbf{Z}$, are modeled by i.i.d. random variables with distribution $\mathcal{CN}\left(0,\sigma_{\mathrm{BS}}^2\right)$. To facilitate the investigation of channel estimation and downlink transmission, we assume that long-term power control is performed to compensate the different strongest AoAs' path loss among different desired users in the desired cell. As a result, it can be considered that large-scale propagation path losses of different users in the desired cell are identical. Thus, we can express the estimated equivalent downlink channel $\widehat{\mathbf{H}}_{\mathrm{eq}}^T \in \mathbb{C}^{N\times N}$ at the desired BS under the impact of pilot contamination in Equation (13) at the top of next page, where $\Delta\widehat{\mathbf{H}}_{\mathrm{eq}}^T$ is the equivalent channel estimation error caused by pilot contamination and noise, and the path loss compensation matrix $\mathbf{B} \in \mathbb{C}^{N\times N}$ is given by

$$\mathbf{B} = \left[\begin{array}{ccc} \frac{1}{\sqrt{\varpi_1}} & \cdots & 0 \\ \vdots & \ddots & \vdots \\ 0 & \cdots & \frac{1}{\sqrt{\varpi_N}} \end{array}\right]. \tag{14}$$

In the following, for notational simplicity, we denote $\widehat{\rho}_{l,k} = \sqrt{\frac{\widehat{\varpi}_{l,k}}{\varpi_k}}$ as the inter-cell propagation path loss coefficients. Now, to evaluate the impact of pilot contamination, we introduce a theorem which reveals the normalized MSE performance of equivalent channel estimation.

*Theorem 1:* The normalized MSE of the equivalent channel estimation with respect to the $k$-th RF chain under the impacts of pilot contamination and noise can be approximated as

$$\begin{aligned}\mathrm{NMSE}_{\mathrm{eq},k} =& \frac{1}{N}\mathbb{E}_{\mathbf{U}_{\mathrm{l,i}}}\left[\left(\frac{1}{\sqrt{MP}}\Delta\widehat{\mathbf{h}}_{\mathrm{eq},k}^T\right)\left(\frac{1}{\sqrt{MP}}\Delta\widehat{\mathbf{h}}_{\mathrm{eq},k}^*\right)\right] \\ &\underbrace{\frac{1}{MP}\sum_{l=1}^{L}\left(\widehat{\rho}_{l,k}^2\right)}_{\text{Multi-cell pilot contamination}} + \underbrace{\frac{\sigma_{\mathrm{BS}}^2}{\varpi_k E_{\mathrm{P}} MP}}_{\text{Noise}}.\end{aligned} \tag{15}$$

In particular, when the number of antennas equipped at the desired BS and the users are sufficiently large, we have

$$\lim_{M,P\to\infty}\mathrm{NMSE}_{\mathrm{eq},k} \approx 0. \tag{16}$$

*Proof:* Please refer to Appendix A. ∎

In Equation (15), the impact of the multi-cell pilot contamination term on the normalized MSE performance is inversely proportional to the number of antennas $M$ and $P$. In addition, the noise term decreases with the increasing transmit pilot symbol energy and the number of antennas, $M$ and $P$. It is important to note that, the impact of noise on channel estimation will vanish in the high signal-to-noise ratio (SNR) regime, e.g. $E_p \gg 1$. However, the impact of pilot contamination on the MSE performance cannot be mitigated by simply increasing the transmit pilot symbol energy $E_p$.

It is known that the conventional massive MIMO pilot-aided LS channel estimation performance under the impact of pilot contamination cannot be improved by increasing the number of antennas equipped at the BS [17], [46]. Interestingly, the result of Theorem 1 unveils that the impacts of pilot contamination and noise on the equivalent channel estimation will vanish asymptotically with the increasing number of antennas equipped at each RF chain, $M$ and $P$. Actually, the numbers of antennas $M$ and $P$ have an identical effect on the normalized MSE performance. This is because the direction of analog beamforming matrices adopted at the desired BS and the desired users align with the strongest AoA path. Hence, the analog beamforming matrices adopted at the desired BS and the users form a pair of spatial filters, which block the pilot signals from undesired users to the desired BS via non-strongest paths. In addition, transmitting the pilot signals from the desired users via the analog beamforming matrix can reduce the potential energy leakage to other undesired cells, which further reduces the impact of pilot contamination.

In order to help the readers to understand Theorem 1 from the physical point of view, we provide a detail discussion in this subsection on the result of Equation (15) via an illustrative example of a single-antenna user in the high SNR regime. In the single-antenna user scenario, we focus on the normalized estimated equivalent channel of user $k$, $\widehat{\mathbf{h}}_{\mathrm{eq},k}^T \in \mathbb{C}^{1\times N}$, in the high SNR regime which is given by

$$\begin{aligned}\frac{1}{\sqrt{M}}\widehat{\mathbf{h}}_{\mathrm{eq},k}^T =& \frac{1}{\sqrt{M}}\mathbf{h}_k^T\underbrace{\left[\begin{array}{ccc} \widehat{\boldsymbol{\nu}}_1 & \ldots & \widehat{\boldsymbol{\nu}}_N \end{array}\right]}_{\mathbf{F}_{\mathrm{RF}}} \\ &+\underbrace{\frac{1}{\sqrt{M}}\sum_{l=1}^{L}\mathbf{u}_{l,k}^T\left[\begin{array}{ccc} \widehat{\boldsymbol{\nu}}_1 & \ldots & \widehat{\boldsymbol{\nu}}_N \end{array}\right]}_{\text{Pilot contamination}}.\end{aligned} \tag{17}$$

From Equation (17), it is shown that the equivalent channel $\widehat{\mathbf{h}}_{\mathrm{eq},k}^T$ is formed by projecting the actual mmWave channel $\mathbf{h}_k$ on the analog beamforming matrix $\mathbf{F}_{\mathrm{RF}}$. Then, $\eta_{k,i} \in \mathbb{C}^{1\times 1}$, the $i$-th entry of the estimated equivalent channel vector $\widehat{\mathbf{h}}_{\mathrm{eq},k}^T$, representing the projection of channel of user $k$ in the desired cell on the $i$-th strongest AoA direction, is given by Equation (18) at the top of next page, where $\theta_{l,k}$ is the AoA from user $k$ in the $l$-th cell to the desired BS, $\mathbf{h}_{\mathrm{S},k}^T$ is the scattering component of the channel of user $k$ in the desired cell, and

$$\widehat{\mathbf{H}}_{\text{eq}}^T = \mathbf{B}\left(\mathbf{H}_{\text{eq}}^T + \Delta\widehat{\mathbf{H}}_{\text{eq}}^T\right) = \mathbf{B}\left(\underbrace{\begin{bmatrix} \widehat{\boldsymbol{\omega}}_1^H \mathbf{H}_1^T \mathbf{F}_{\text{RF}} \\ \vdots \\ \widehat{\boldsymbol{\omega}}_N^H \mathbf{H}_N^T \mathbf{F}_{\text{RF}} \end{bmatrix}}_{\mathbf{H}_{\text{eq}}^T} + \underbrace{\frac{1}{\sqrt{E_{\text{P}}}}\begin{bmatrix} \boldsymbol{\Phi}_1^H \mathbf{Z}^T \mathbf{F}_{\text{RF}} \\ \vdots \\ \boldsymbol{\Phi}_N^H \mathbf{Z}^T \mathbf{F}_{\text{RF}} \end{bmatrix}}_{\text{Effective noise}} + \underbrace{\begin{bmatrix} \sum_{l=1}^{L}\left(\widehat{\boldsymbol{\omega}}_{l,1}^H \mathbf{U}_{l,1}^T\right)\mathbf{F}_{\text{RF}} \\ \vdots \\ \sum_{l=1}^{L}\left(\widehat{\boldsymbol{\omega}}_{l,N}^H \mathbf{U}_{l,N}^T\right)\mathbf{F}_{\text{RF}} \end{bmatrix}}_{\text{Pilot contamination}}\right). \tag{13}$$

$$\eta_{k,i} = \frac{1}{\sqrt{M}}\mathbf{h}_k^T\widehat{\boldsymbol{\nu}}_i + \frac{1}{\sqrt{M}}\sum_{l=1}^{L}\mathbf{u}_{l,k}^T\widehat{\boldsymbol{\nu}}_i = \underbrace{\sqrt{\frac{\varsigma_k}{\varsigma_k+1}}\frac{\sin\left[M\pi\frac{d}{\lambda}\left(\cos\left(\theta_k\right)-\cos\left(\theta_i\right)\right)\right]}{M\sin\left[\pi\frac{d}{\lambda}\left(\cos\left(\theta_k\right)-\cos\left(\theta_i\right)\right)\right]}}_{\text{Strongest AoA}} + \underbrace{\sqrt{\frac{1}{\varsigma_k+1}}\frac{1}{\sqrt{M}}\mathbf{h}_{\text{S},k}^T\widehat{\boldsymbol{\nu}}_i}_{\text{Scattering component}}$$
$$+\underbrace{\sum_{l=1}^{L}\sqrt{\frac{\varsigma_{l,k}}{\varsigma_{l,k}+1}}\frac{\sin\left[M\pi\frac{d}{\lambda}\left(\cos\left(\theta_{l,k}\right)-\cos\left(\theta_i\right)\right)\right]}{M\sin\left[\pi\frac{d}{\lambda}\left(\cos\left(\theta_{l,k}\right)-\cos\left(\theta_i\right)\right)\right]} + \sum_{l=1}^{L}\sqrt{\frac{1}{\varsigma_{l,k}+1}}\frac{1}{\sqrt{M}}\mathbf{u}_{\text{S},l,k}^T\widehat{\boldsymbol{\nu}}_i}_{\text{Pilot contamination}}. \tag{18}$$

$\mathbf{u}_{\text{S},l,k}^T$ is the scattering component of the channel from user $k$ in the $l$-th cell to the desired BS. For the projection of channel of user $k$ on the $k$-th RF chain's analog beamforming vector $\widehat{\boldsymbol{\nu}}_k$, we have $\widehat{\boldsymbol{\nu}}_k^T = \left(\mathbf{h}_{\text{SAoA},k}^{\text{BS}}\right)^H$. Thus, the $k$-th entry of the vector $\widehat{\mathbf{h}}_{\text{eq},k}^T$ can be expressed as

$$\eta_{k,k} = \underbrace{\sqrt{\frac{\varsigma_k}{\varsigma_k+1}}}_{\text{Strongest AoA}} + \underbrace{\varepsilon_k\frac{1}{\sqrt{M}}\sqrt{\frac{1}{\varsigma_k+1}}}_{\text{Scattering component}} + \underbrace{\frac{1}{\sqrt{M}}\mu_k}_{\text{Pilot contamination}}, \tag{19}$$

where $\varepsilon_k = \mathbf{h}_{\text{S},k}^T\widehat{\boldsymbol{\nu}}_k$ denotes the projection of the scattering component $\sqrt{\frac{1}{\varsigma_k+1}}\mathbf{h}_{\text{S},k}$ on $\widehat{\boldsymbol{\nu}}_k$ and $\mu_k = \sum_{l=1}^{L}\mathbf{u}_{l,k}^T\widehat{\boldsymbol{\nu}}_k$ denotes the projection of inter-cell channels $\sum_{l=1}^{L}\mathbf{u}_{l,k}^T$ on $\widehat{\boldsymbol{\nu}}_k$.

The computation of Equation (19) can be illustrated graphically via the concept of vector space with projection as shown in Fig. 2(a). The first term of Equation (19), $\sqrt{\frac{\varsigma_k}{\varsigma_k+1}}$, which is the projection of the strongest AoA component on the analog beamforming direction, is a constant and is independent of $M$. For the second term and the third term of Equation (19), $\varepsilon_k$ and $\mu_k$, which consist of multiple paths with various AoAs, cannot enjoy the array gain $M$. Actually, $\frac{\varepsilon_k}{\sqrt{M}}\sqrt{\frac{1}{\varsigma_k+1}}$ and $\frac{\mu_k}{\sqrt{M}}$, these two terms will vanish asymptotically with the increasing number of antennas $M$.

We note that for the conventional pilot-aided channel estimation algorithms, e.g. LS-based algorithms, they estimate the channels from all the directions. Thus, BSs adopting these algorithms receiving reused pilot symbols from the undesired users and cannot be distinguished from the desired pilot symbols. It is known as pilot contamination. However, adopting analog beamforming matrices for receiving pilot symbols at the desired BS via the strongest AoA directions forms a spatial filter, which blocks the undesired pilot symbols from neighboring cells via different AoA paths. Furthermore, the "blocking capability" improves with the increasing number of antennas equipped at the BS. Specifically, the beamwidth of mainlobe becomes narrower and the magnitude of sidelobes is lower, which is illustrated in Fig. 2(b). In fact, this is an important feature for mitigating the impact of pilot contamination.

Therefore, with the equivalent channel estimation proposed in [2], the impact of pilot contamination vanishes asymptotically with an increasing number of antennas $M$ equipped at the desired BS, even if there exists errors in the estimation of the strongest AoAs[10].

## IV. Multi-cell Downlink Achievable Rate

In the last section, we mathematically prove that the impact of pilot contamination on the equivalent channel estimation can be effectively mitigated. Analytical and simulation results show that CSI errors caused by pilot contamination vanish asymptotically with the increasing numbers of antennas equipped at each RF chain of the BS and the users, $M$ and $P$.

Now, we aim to study the downlink performance of multi-cell hybrid mmWave networks, which takes into account the impact of inter-cell interference caused by the neighboring BSs as well as the intra-cell interference caused by channel estimation errors. In this section, we illustrate and derive the closed-form approximations of achievable rate per user under the impact of intra-cell and the inter-cell interference of the considered hybrid system. Our results reveal that the intra-cell interference as well as the inter-cell interference can be mitigated effectively by simply increasing the number of antennas equipped at the desired BS.

### A. Downlink Achievable Rate Performance

We now detail the received information signal at user $k$ in the desired cell. Due to the channel reciprocity, we utilize

[10]Generally, analog beamforming errors may cause extra rate performance degradation, which requires more antenna array gains to compensate. Interested readers can refer to Equations (31)-(34), Theorem 2, Corollary 4, and Fig. 7 in [2] for more details.

$$\mathbb{E}_{\Delta\widehat{\mathbf{H}}_{\mathrm{eq}}^{T},\mathbf{H}_{l,k}}\left[\widehat{\mathrm{SINR}}_{\mathrm{ZF}}^{k}\right] \approx \frac{\overline{\beta}^2 \varpi_k E_s}{\underbrace{\overline{\beta}^2 \varpi_k E_s \widehat{\mathbf{h}}_{\mathrm{eq},k}^{T} \mathbb{E}_{\Delta\widehat{\mathbf{H}}_{\mathrm{eq}}^{T}}\left[\Delta\overline{\mathbf{W}}_{\mathrm{eq}}\Delta\overline{\mathbf{W}}_{\mathrm{eq}}^{H}\right]\widehat{\mathbf{h}}_{\mathrm{eq},k}^{*}}_{\mathrm{Intra-cell\ \ interference}\ \ \Upsilon_k} + \underbrace{\mathbb{E}_{\mathbf{H}_{l,k}}\left[\left|\widehat{\boldsymbol{\omega}}_k^{H}\sum_{l=1}^{L}\mathbf{H}_{l,k}^{T}\mathbf{F}_{\mathrm{RF},l}\widehat{\beta}_l\widehat{\mathbf{W}}_{\mathrm{eq},l}\mathbf{x}_l\right|^2\right]}_{\mathrm{Inter-cell\ \ interference}\ \ \Omega_k} + \sigma_{\mathrm{MS}}^2}. \tag{22}$$

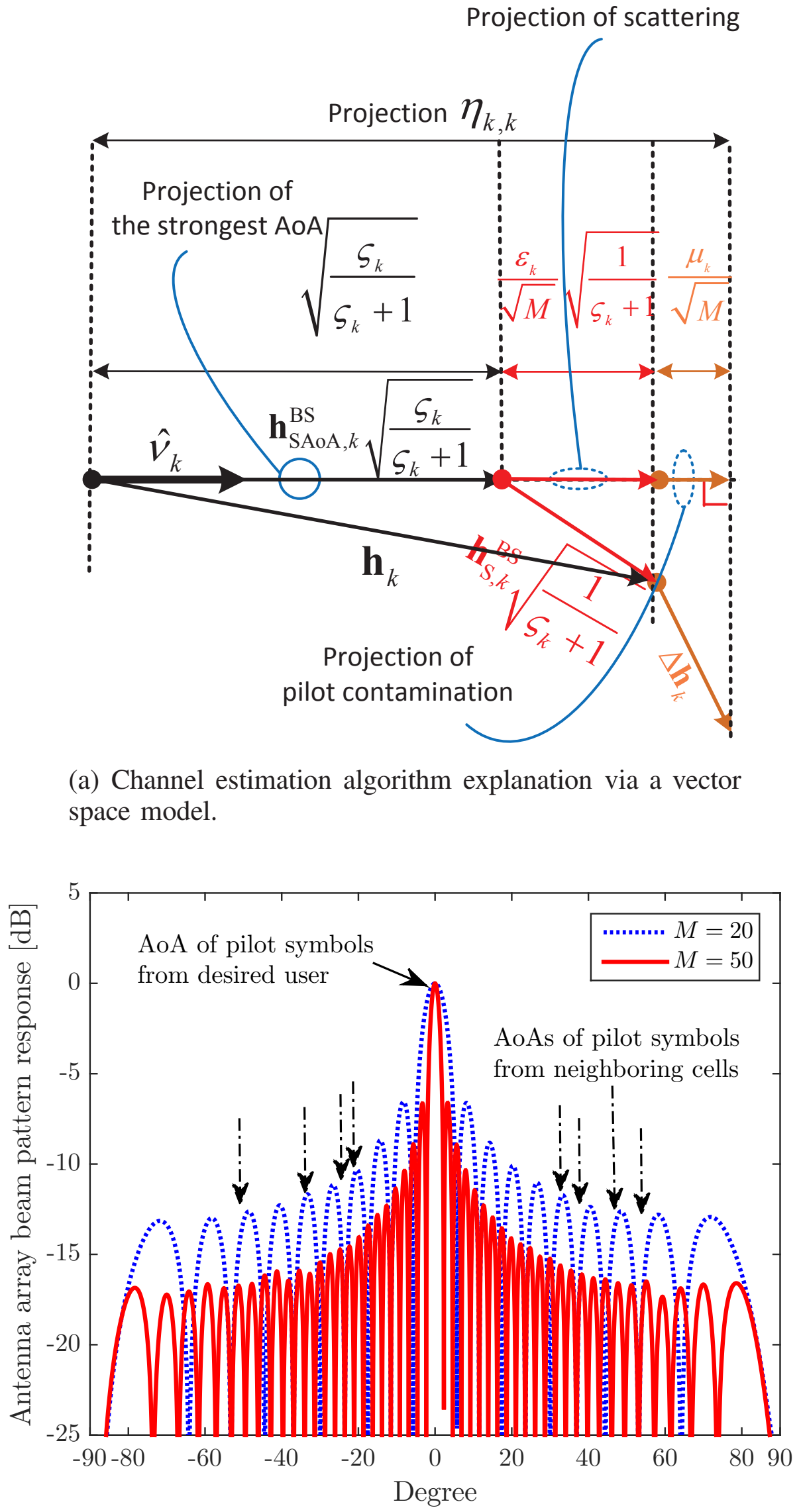

(a) Channel estimation algorithm explanation via a vector space model.

(b) Antenna array beam pattern response.

Fig. 2. (a) The illustration of equivalent channel estimation under the impact of pilot contamination. (b) The illustration of sidelobe suppression for different numbers of antennas $M$.

the estimated equivalent channel $\widehat{\mathbf{H}}_{\mathrm{eq}}^{T}$ in Equation (13) for the digital baseband precoder design during the downlink transmission. To suppress the inter-user interference among different users, we adopt a ZF precoder for the downlink transmission. In addition, equal power allocation is used among different data streams of the users [30]. The desired baseband digital ZF precoder $\overline{\mathbf{W}}_{\mathrm{eq}} \in \mathbb{C}^{N\times N}$ is given by

$$\overline{\mathbf{W}}_{\mathrm{eq}} = \mathbf{H}_{\mathrm{eq}}^{*}(\mathbf{H}_{\mathrm{eq}}^{T}\mathbf{H}_{\mathrm{eq}}^{*})^{-1} = \left[\ \overline{\mathbf{w}}_{\mathrm{eq},1},\ldots,\overline{\mathbf{w}}_{\mathrm{eq},N}\ \right]. \tag{20}$$

The received signal at user $k$ in the desired cell under the intra-cell and inter-cell interference is given by

$$\begin{aligned} y_{\mathrm{ZF}}^{k} =& \underbrace{\widehat{\boldsymbol{\omega}}_k^{H}\mathbf{H}_k^{T}\mathbf{F}_{\mathrm{RF}}\overline{\beta}\overline{\mathbf{w}}_{\mathrm{eq},k}x_k}_{\text{Desired signal}} + \underbrace{\widehat{\boldsymbol{\omega}}_k^{H}\mathbf{H}_k^{T}\mathbf{F}_{\mathrm{RF}}\sum_{j=1}^{N}\overline{\beta}\Delta\overline{\mathbf{w}}_{\mathrm{eq},j}x_j}_{\mathrm{Intra-cell\ \ interference}} \\ &+ \underbrace{\widehat{\boldsymbol{\omega}}_k^{H}\sum_{l=1}^{L}\mathbf{H}_{l,k}^{T}\mathbf{F}_{\mathrm{RF},l}\widehat{\beta}_l\widehat{\mathbf{W}}_{\mathrm{eq},l}\mathbf{x}_l}_{\mathrm{Inter-cell\ \ interference}} + \underbrace{\widehat{\boldsymbol{\omega}}_k^{H}\mathbf{z}_{\mathrm{MS},k}}_{\text{Noise}}, \end{aligned} \tag{21}$$

where $x_k \in \mathbb{C}^{1\times 1}$ is the transmitted symbol from the desired BS to the desired user $k$, $\mathbb{E}\left[\left|x_k^2\right|\right] = E_s$ is the average transmitted symbol energy for each user, $\overline{\beta} = \sqrt{\frac{1}{\mathrm{tr}(\overline{\mathbf{W}}_{\mathrm{eq}}\overline{\mathbf{W}}_{\mathrm{eq}}^{H})}}$ is the transmission power normalization factor of the desired BS, $\Delta\overline{\mathbf{w}}_{\mathrm{eq},j} \in \mathbb{C}^{N\times 1}$ denotes the $j$-th column vector of the ZF precoder error matrix $\Delta\overline{\mathbf{W}}_{\mathrm{eq}} = \widehat{\overline{\mathbf{W}}}_{\mathrm{eq}} - \overline{\mathbf{W}}_{\mathrm{eq}} = \left[\ \Delta\overline{\mathbf{w}}_{\mathrm{eq},1},\ldots,\Delta\overline{\mathbf{w}}_{\mathrm{eq},N}\ \right]$, $\widehat{\overline{\mathbf{W}}}_{\mathrm{eq}} = \widehat{\mathbf{H}}_{\mathrm{eq}}^{*}(\widehat{\mathbf{H}}_{\mathrm{eq}}^{T}\widehat{\mathbf{H}}_{\mathrm{eq}}^{*})^{-1}$ is the designed ZF precoder based on the estimated equivalent channel, and the effective noise $\mathbf{z}_{\mathrm{MS},k} \sim \mathcal{CN}\left(\mathbf{0},\sigma_{\mathrm{MS}}^2\mathbf{I}\right)$. In addition, $\widehat{\mathbf{W}}_{\mathrm{eq},l} = \widehat{\mathbf{H}}_{\mathrm{eq},l}^{*}(\widehat{\mathbf{H}}_{\mathrm{eq},l}^{T}\widehat{\mathbf{H}}_{\mathrm{eq},l}^{*})^{-1}$ is the digital precoder of the BS in cell $l$ based on the estimated equivalent channel $\widehat{\mathbf{H}}_{\mathrm{eq},l}$, $\mathbf{F}_{\mathrm{RF},l}$ is the analog beamforming matrix of the BS in cell $l$, $\mathbf{x}_l = [x_{l,1}, x_{l,2},\ldots, x_{l,N}]^T$ denotes the transmitted signal for all the users in cell $l$, and $\widehat{\beta}_l = \sqrt{\frac{1}{\mathrm{tr}\left(\widehat{\mathbf{W}}_{\mathrm{eq},l}\widehat{\mathbf{W}}_{\mathrm{eq},l}^{H}\right)}}$ is the power normalization factor of the BS in cell $l$. Compared to the system model in [2], the intra-cell interference, which is caused by CSI errors due to pilot contamination effects, is captured and investigated. In addition, the model considered here also captures the inter-cell interference caused by the simultaneous transmission from the neighboring BSs.

We then express the average receive signal-to-interference-plus-noise ratio (SINR) expression of user $k$ in the desired cell as Equation (22) at the top of this page.

### B. Intra-Cell and Inter-cell Interference Performance Analysis

In this section, we aim at deriving a closed-form expression of the average achievable rate per user. To this end, we first focus on the intra-cell interference term $\Upsilon_k$ shown in Equation (22). Note that the intra-cell interference $\Upsilon_k$ is caused by CSI errors, which is due to the impact of pilot contamination on channel estimation. Then we detail and analyze the inter-cell interference $\Omega_k$. The expression of intra-cell interference, $\Upsilon_k$, can be summarized in the following lemma.

*Lemma 1:* In the large number of antennas regimes, the asymptotic intra-cell interference under the impact of pilot contamination can be approximated by

$$\Upsilon_k = \overline{\beta}^2 \varpi_k E_s \widehat{\mathbf{h}}_{\mathrm{eq},k}^T \mathbb{E}_{\Delta\widehat{\mathbf{H}}_{\mathrm{eq}}^T}\left[\Delta\overline{\mathbf{W}}_{\mathrm{eq}}\Delta\overline{\mathbf{W}}_{\mathrm{eq}}^H\right]\widehat{\mathbf{h}}_{\mathrm{eq},k}^* \overset{M\to\infty}{\approx} \overline{\beta}^2 \varpi_k E_s \left[\left(\sqrt{1+\sum_{l=1}^{L}\frac{\widehat{\rho}_{l,k}^2}{MP}}-1\right)^2 + \left(1+\sum_{l=1}^{L}\frac{\widehat{\rho}_{l,k}^2}{MP}\right)\sum_{l=1}^{L}\frac{\widehat{\rho}_{l,k}^2 N}{MP}\frac{\varsigma_k+1}{\varsigma_k}\right]. \quad (23)$$

*Proof:* The proof follows a similar approach as in [2]. Due to the page limitation, we omit the proof here. ■

Now, we introduce the following lemma which simplifies the expression of the intra-cell interference.

*Lemma 2:* In the large numbers of antennas regime, the asymptotic inter-cell interference, which is caused by the BSs in neighboring cells, can be approximated as

$$\Omega_k = \mathbb{E}_{\mathbf{H}_{l,k}}\left[\left|\widehat{\boldsymbol{\omega}}_k^H \sum_{l=1}^{L}\mathbf{H}_{l,k}^T \mathbf{F}_{\mathrm{RF},l}\widehat{\beta}_l \widehat{\mathbf{W}}_{\mathrm{eq},l}\mathbf{x}_l\right|^2\right] \overset{M\to\infty}{\approx} E_s\left[\sum_{l=1}^{L}\left(\widetilde{\varpi}_{l,k}\right)\right]. \quad (24)$$

*Proof:* Please refer to Appendix B. ■

From Equation (24), the inter-cell interference is mainly contributed by the downlink transmission of neighboring BSs. Besides, the power normalization factor $\overline{\beta}$ can be approximated as [2]

$$\overline{\beta} \approx \sqrt{\frac{\varsigma_k}{\varsigma_k+1}\frac{MP}{N}}. \quad (25)$$

Then, we summarize the average achievable rate of user $k$ in the large number of antennas and high SNR regime in the following theorem.

*Theorem 2:* For a large number of antennas regime, the average achievable rate of user $k$ adopting the ZF precoding under the intra-cell and the inter-cell interference is approximated by

$$\widetilde{R}_{\mathrm{ZF}}^k \approx \log_2\left\{1+\left[\left(\sqrt{1+\frac{1}{MP}\sum_{l=1}^{L}\left(\widehat{\rho}_{l,k}^2\right)}-1\right)^2 + \left[1+\frac{1}{MP}\sum_{l=1}^{L}\left(\widehat{\rho}_{l,k}^2\right)\right]\frac{N\left(\varsigma_k+1\right)}{\varsigma_k MP}\sum_{l=1}^{L}\left(\widehat{\rho}_{l,k}^2\right) + \frac{N(\varsigma_k+1)}{\varsigma_k MP}\sum_{l=1}^{L}\left(\zeta_{l,k}^2\right)+\frac{\sigma_{\mathrm{MS}}^2}{\overline{\beta}^2\varpi_k E_s}\right]^{-1}\right\}, \quad (26)$$

where $\zeta_{l,k} = \sqrt{\frac{\widetilde{\varpi}_{l,k}}{\varpi_k}}$.

*Proof:* The result follows by substituting Equations (23), (24), and (25) into (22). ■

In the high SNR and large antenna regimes, the average achievable rate under the impact of inter-cell interference caused by CSI errors and the intra-cell interference caused by neighboring BSs can be mitigated by simply increasing the number of antennas $M$. Thus, the average achievable rate per user can increase unboundedly with an increasing antenna $\log_2 M$.

For comparison, we also derive the achievable rate of user $k$ in a single-cell hybrid system. The performance of this system, $R_{\mathrm{HB}}^{\mathrm{upper}_k}$, serves as a performance upper bound of the considered multi-cell system as there is no out-of-cell interference. The expression of $R_{\mathrm{HB}}^{\mathrm{upper}_k}$ is given by (Equation (24) of Corollary 1 in [2])

$$R_{\mathrm{HB}}^{\mathrm{upper}_k} \overset{a.s.}{\underset{M\to\infty}{\to}} \log_2\left\{1+\left[\frac{MP}{N}\frac{\varsigma_k}{\varsigma_k+1}+\frac{1}{\varsigma_k+1}\right]\frac{E_s}{\sigma_{\mathrm{MS}}^2}\right\}. \quad (27)$$

*Corollary 1:* Now, we derive the scaling law of the average achievable rate of user $k$ in the high SNR regime which is given by

$$\lim_{M\to\infty}\frac{\widetilde{R}_{\mathrm{ZF}}^k}{\log_2 M} = \lim_{\mathrm{M}\to\infty}\frac{R_{\mathrm{HB}}^{\mathrm{upper}_k}}{\log_2 M} = 1. \quad (28)$$

*Proof:* The result comes after some straightforward mathematical manipulation. Due to the page limitation, we omit the proof here. ■

It is found that the system data rate scales with $\log_2 M$. In other words, pilot contamination is not a fundamental problem for massive MIMO with hybrid mmWave systems, when the CSI estimation algorithm proposed in [2] is adopted. In contrast, the average achievable rate performance of multi-cell MU massive MIMO networks adopting a conventional LS-based channel estimation algorithm is limited by the effects of pilot contamination, e.g. [11]. Note that the algorithm proposed in [2] does not require any information of covariance matrices of pilot-sharing users in neighbouring cells as required by the multi-cell MMSE-based precoding algorithm proposed in [54]. In addition, the algorithm proposed in [2] can work well while mean values of cross-cell channels from pilot-sharing users in neighbouring cells to the desired BS are non-zero.

*Remark 1:* Since hybrid mmWave systems are the generalization of fully digital systems, the algorithm proposed in [2] can be extended to the case of fully digital mmWave systems. Under such circumstance, the derived analysis can be directly applied to the latter systems.

## V. Discussions and Simulations

In this section, we will discuss and verify the derived results via simulations. In the following simulations unless specified otherwise, simulation settings are listed as follows. The transmit antenna gain of the BS antennas is assumed as 14 dBi. The maximum BS transmit power is set as 46 dBm. We set the number of neighboring cells as $L = 6$ and the number of users per cell $N = 10$. For the adopted simulation parameters, the large-scale path loss coefficients[11], $\alpha$ and $\varrho$, may have different values for different scenarios. We have $\alpha = 1.9$ and $\varrho = 20$ for corresponding large-scale path loss calculation [35]. In addition, the receiving thermal noise power is $\kappa = -90$ dBm[12]. The channel estimation errors of a user caused by pilot contamination is given by $\xi^2 = \sum_{l=1}^{L} \left(\widehat{\rho}_{l,k}^2\right)$, $k \in \{1, \ldots, N\}$. Besides, we define the average achievable rate per user as $\widetilde{R}_{\mathrm{AVE}} = \frac{1}{N} \sum_{k=1}^{N} \widetilde{R}_{\mathrm{ZF}}^k$.

[11]The commonly adopted empirical propagation path loss model is considered as a function of distance and carrier frequency, which is given by $\varpi_k = 10^{-\left(\frac{10\alpha\log_{10} d_k+\varrho\log_{10}\left(4\pi\frac{1}{\lambda}\right)}{10}\right)}$, where $d_k$ is the distance between the desired BS and user $k$ in the desired cell, $\alpha$ and $\varrho$ are the least square fits of floating intercept and slope over the measured distance [35].

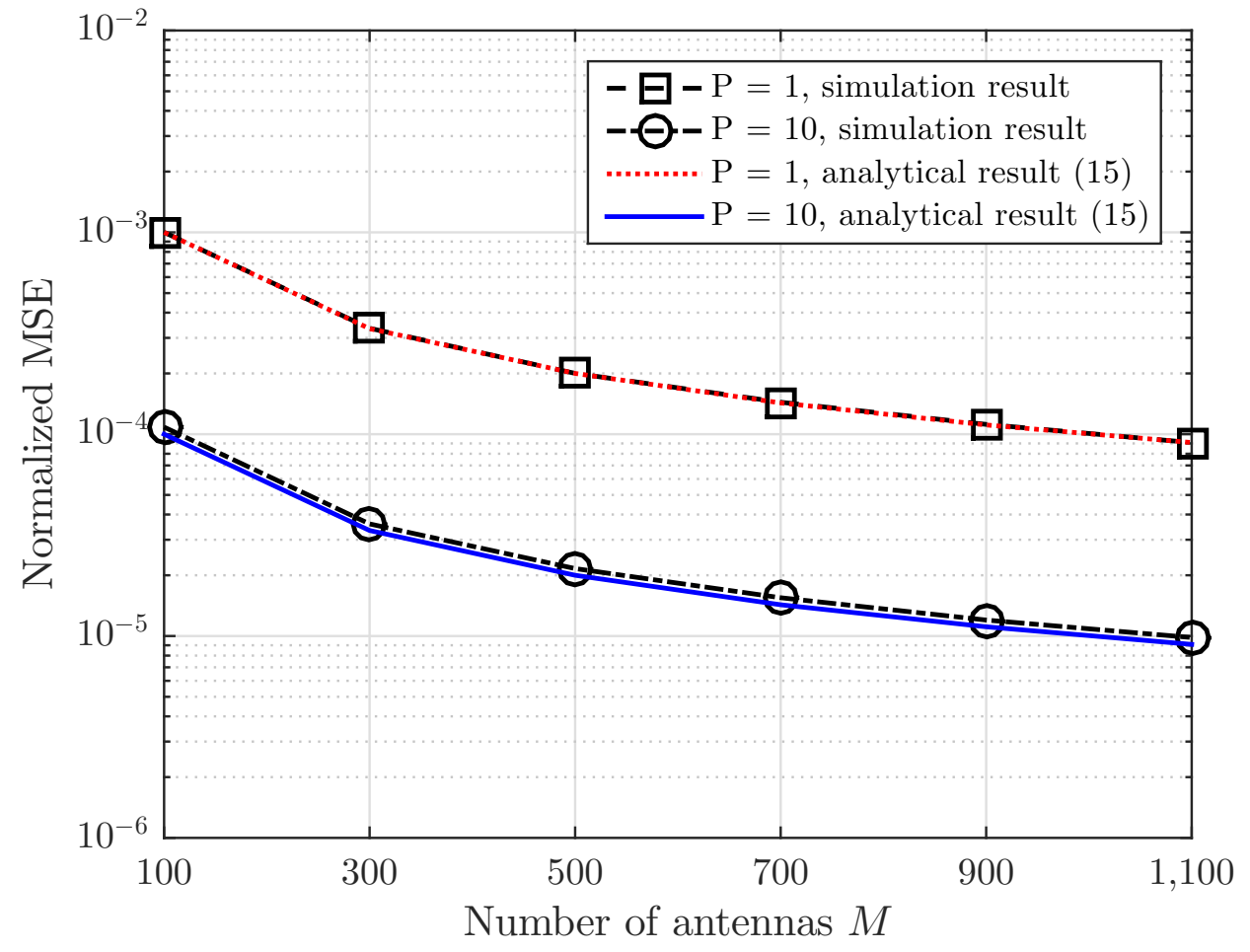

Fig. 3. The illustration of multi-cell normalized MSE performance under the impact of pilot contamination versus the number of antennas $M$ in the high SNR regime, i.e., maximum transmit power 46 dBm, for $N_{\mathrm{RF}} = N = 10$, $\varsigma_{l,k} = 5$, $k \in \{1, \cdots, N\}$ and channel estimation error $\xi^2 = 0.01$.

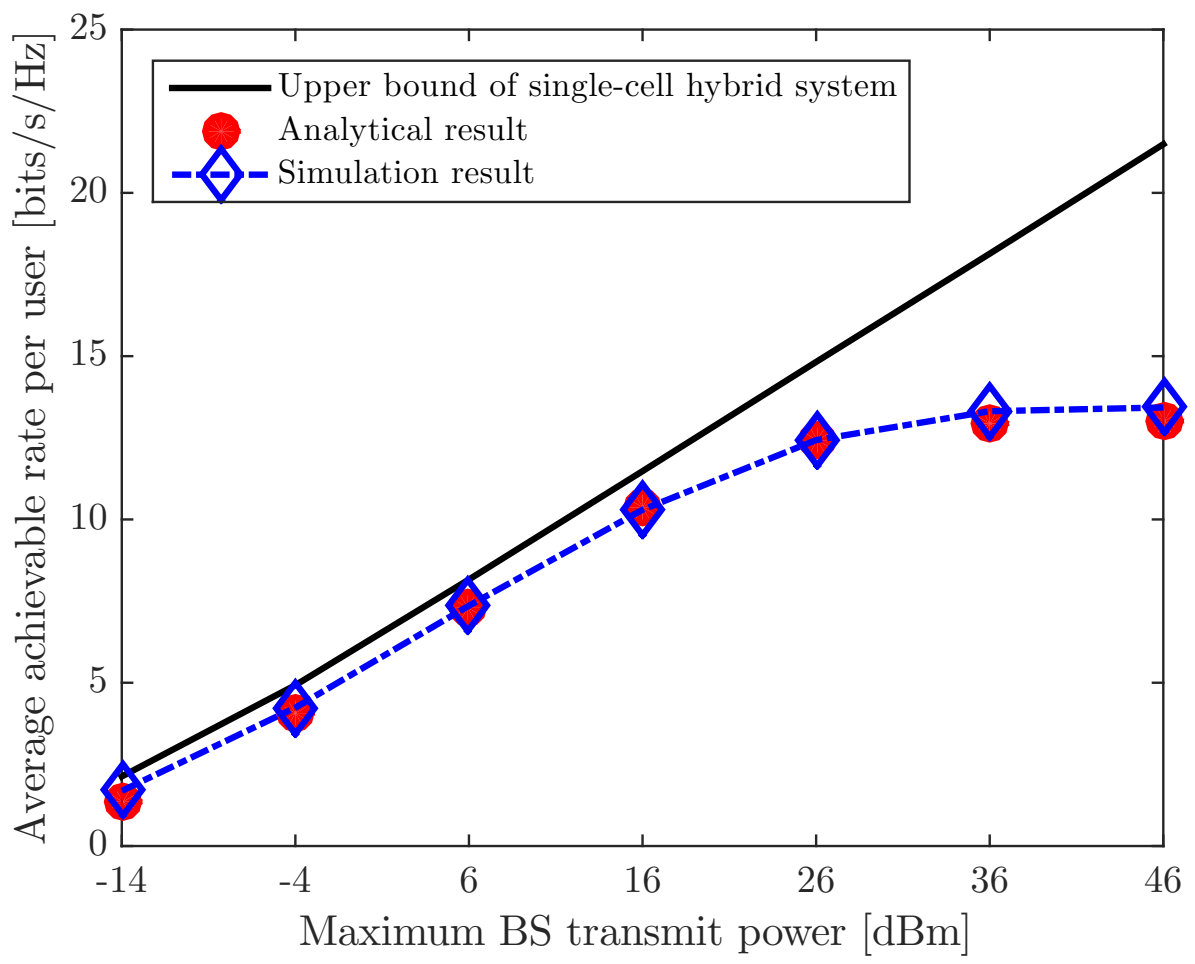

Fig. 4. The average achievable rate per user [bits/s/Hz] under the impact of pilot contamination and inter-cell downlink transmission interference versus maximum transmit power at the desired BS [dBm] with $M = 200$.

To verify the correctness of the analytical results derived in Equation (15), here, we provide some simulation results in Fig. 3, which illustrates the normalized MSE performance of the equivalent channel estimation versus the number of antennas equipped at the desired BS, $M$, under the impact of pilot contamination. In the simulation, we take into account the inter-cell propagation path loss and AoA estimation errors in estimating the strongest AoA paths. For illustration, channel estimation errors caused by pilot contamination is set as $\xi^2 = 0.01$. Due to the existence of the inter-cell strongest AoA components, this setting can be considered as the worst case scenario of pilot contamination as the impact of inter-cell interference is magnified. In Fig. 3, we can observe that with an increasing number of antennas $M$, the normalized MSE decreases monotonically. In addition, an increasing number of antennas equipped at the users $P$ can also improve the normalized MSE. Besides, the simulation results match the analytical results derived in Equation (15). Thus, the impact of pilot contamination on the channel estimation of multi-cell hybrid mmWave systems vanishes asymptotically, for a sufficiently large number of antennas equipped at the desired BS $M$.

In Fig. 4, we verify the tightness of the derived approximation in Equation (26) of Theorem 2. Fig. 4 illustrates the average achievable rate per user under the intra-cell and inter-cell interference versus SNR. Besides, we illustrate the corresponding upper performance which is based on a single-cell hybrid mmWave system. In Fig. 4, the number of antennas equipped at the users is set to $P = 10$, intra-cell $\varsigma_k = 4$, inter-cell $\varsigma_{l,k} = 2$, $k \in \{1, \ldots, N\}$, the channel estimation error caused by pilot contamination is set as $\xi^2 = 0.01$. We can observe that the derived approximation is tight for a wide range of SNR. It also can be observed that, with a finite number of antennas, there is still a ceiling of the achievable rate performance in the high SNR regime.

The setup in Fig. 5 is considered at intra-cell $\varsigma_k = 4$, inter-cell $\varsigma_{l,k} = 2$, $k \in \{1, \ldots, N\}$, and the number of antennas equipped at each user is $P = 10$. In Fig. 5, we first illustrate the upper bounds of the average achievable rate per user of the single-cell hybrid mmWave system and the single-cell fully digital system as the system benchmarks. Without any intra-cell interference and inter-cell interference, the corresponding upper performance of the fully digital system as well as the hybrid mmWave system increases linearly with the increasing number of BS antennas $M$. Then, we compare the achievable rate per user of the fully digital system using a conventional LS-based CSI estimation algorithm to that of the hybrid system adopting CSI estimation algorithm proposed in [2] in the high SNR regime. For the fully digital system, we assume that the receive CSI at the users' side is perfectly known. Thus, the $P$ antenna array equipped at each user can provide $10\log_{10} P$ dB array gain. In addition, we assume that channel estimation errors of a user at the desired BS due to the impact of pilot contamination effects for both the fully digital system and the hybrid system are set as $\xi^2 = 0.2$. As expected, in the large number of antennas and high SNR regimes, the average achievable rate of the hybrid system increases unboundedly and logarithmically with the increasing number of antennas $M$, despite the existence of pilot contamination as well as inter-cell interference. The average achievable rate of the hybrid system scales with the number of BS antennas in the order of $\log_2 M$, which is the same as that of the upper performance of the single-cell hybrid system, and verifies the correctness of the derived scaling law (Equation (28) in Corollary 1). However, the average achievable rate of the fully digital system adopting the conventional LS-based CSI estimation algorithm under the impact of pilot contamination is saturated with the increasing number of antennas at the desired BS. Thus, pilot contamination is the performance bottleneck of massive MIMO systems when the conventional LS-based channel estimation algorithm is adopted [11]. In contrast, when the channel estimation algorithm proposed in [2] is applied to the multi-cell MU mmWave hybrid system, the impacts of CSI errors caused by pilot contamination and the inter-cell interference caused by neighboring BSs on the achievable rate performance will vanish asymptotically for a sufficiently large number of antennas equipped at the desired BS. For a practical number of antennas setup, $M \in [64,\ 192]$, the average rate performance achieved by the algorithm proposed in [2] can also increases with an increasing number of antennas $M$ and outperforms that of a fully digital system, which is illustrated in Fig. 5(b). Meanwhile, the number of RF chains required in the hybrid system is significantly smaller than that of fully digital system. In addition, our simulation results in Fig. 5(a) also verify the tightness of the analytical approximation derived in Theorem 2.

[12]Thermal noise power is determined by the signal bandwidth $B_W$ and the noise spectral density level $N_0 = B_{\mathrm{W}} e_{\mathrm{B}} T_k$, where $B_{\mathrm{W}} = 250$ MHz, $T_k = 300$ K, and $e_{\mathrm{B}} = 1.38 \times 10^{-23}$ J/K is the Boltzmann constant [35].

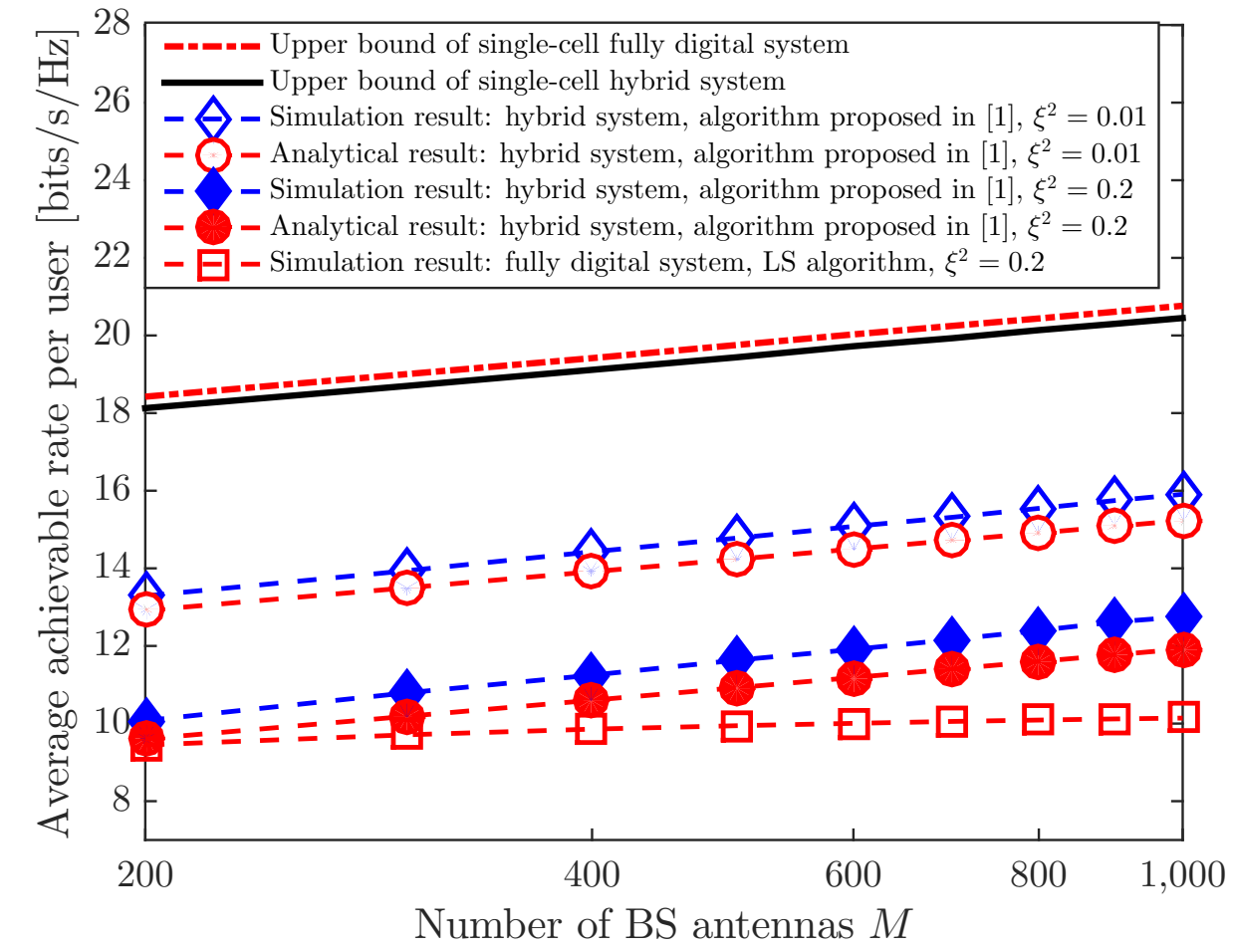

(a)

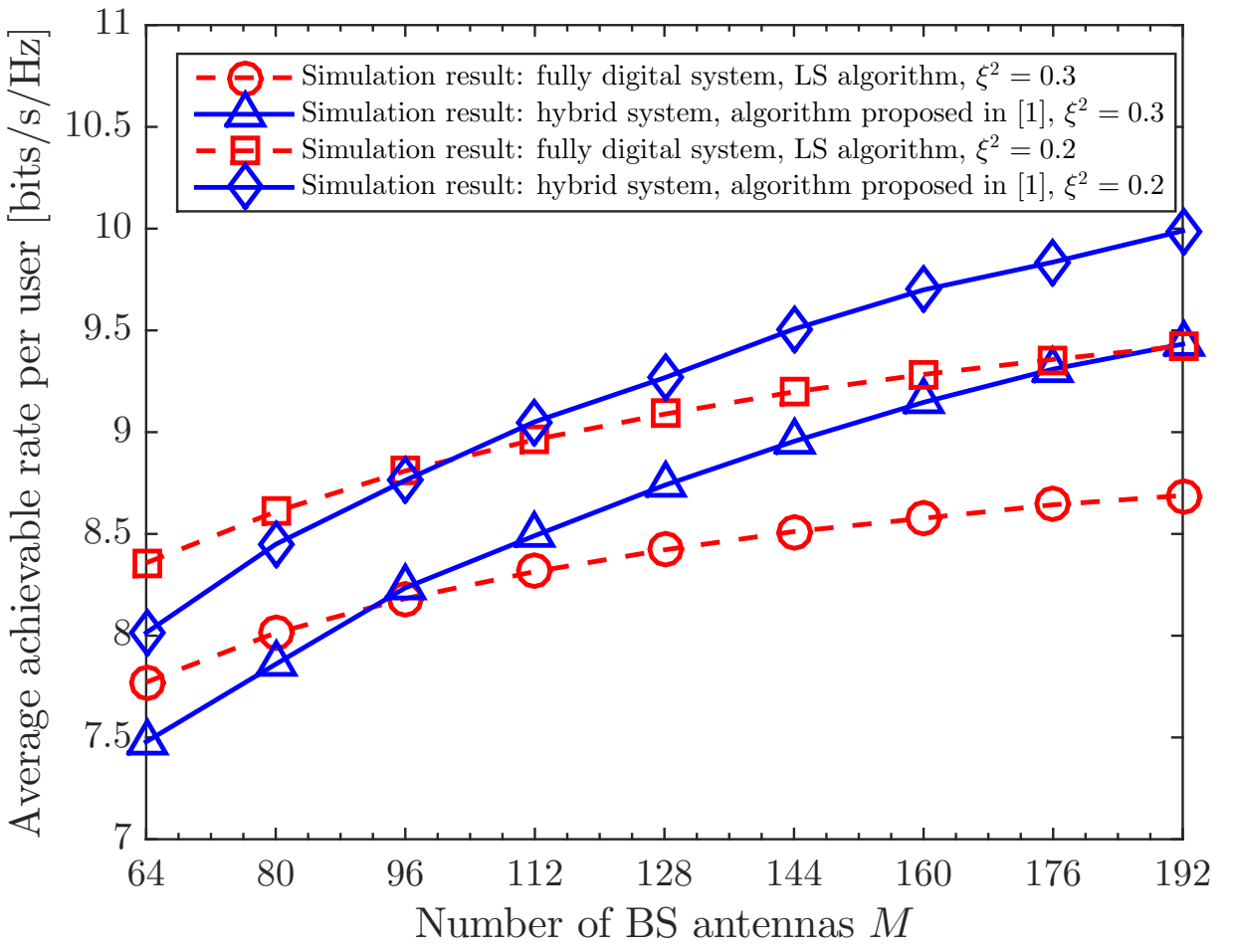

(b)

Fig. 5. The achievable rate per user [bits/s/Hz] under the impact of pilot contamination effects and inter-cell interference versus the number of antennas $M$ in the high SNR regime (maximum BS transmit power is 46 [dBm]).

## VI. Conclusions

In this paper, we investigated the normalized MSE performance of the channel estimation proposed in [2] for multi-cell hybrid mmWave systems. The derived closed-form approximation of the normalized channel estimation MSE performance revealed that the channel estimation error caused by the impact of pilot contamination and noise would vanish asymptotically with the increasing number of antennas. Furthermore, based on the estimated CSI, we adopted ZF precoding for the downlink transmission and derived the closed-form approximation of the average achievable rate per user in the multi-cell scenario. The analytical and simulation results showed that the intra-cell interference caused by pilot contamination as well as the inter-cell interference incurred by neighboring BSs can be mitigated effectively with an increasing antenna array gain. It is an excellent feature for multi-cell hybrid mmWave systems with small-cell radius for improving the network spectral efficiency.

## Appendix

### A. Proof of Theorem 1

The normalized MSE performance of equivalent channel estimation under the impact of pilot contamination is given by

$$\begin{aligned}\mathrm{NMSE}_{\mathrm{eq},k} &= \frac{1}{NMP}\mathbb{E}_{\mathbf{U}_{l,i}}\left[\left(\Delta\widehat{\mathbf{h}}_{\mathrm{eq},k}^{T}\right)\left(\Delta\widehat{\mathbf{h}}_{\mathrm{eq},k}^{*}\right)\right] \\ &\overset{(a)}{\approx} \underbrace{\frac{1}{\varpi_k NMP}\mathbb{E}_{\mathbf{U}_{l,i}}\left[\sum_{l=1}^{L}\left(\widehat{\boldsymbol{\omega}}_{l,k}^{H}\mathbf{U}_{l,k}^{T}\mathbf{F}_{\mathrm{RF}}\right)\times\sum_{l=1}^{L}\left(\mathbf{F}_{\mathrm{RF}}^{H}\mathbf{U}_{l,k}^{*}\widehat{\boldsymbol{\omega}}_{l,k}\right)\right]}_{\text{Pilot contamination}} \\ &+\underbrace{\frac{\sigma_{\mathrm{BS}}^{2}\mathrm{tr}\left[\mathbf{F}_{\mathrm{RF}}^{H}\mathbf{F}_{\mathrm{RF}}\right]}{\varpi_k E_{\mathrm{P}} NMP}}_{\text{Effective noise}}. \qquad (29)\end{aligned}$$

In (a), we omit some small part. Due to the small-cell radius, the inter-cell uplink propagation channels contain the strongest AoA components from the users in the neighboring cells to the desired BS. Thus, the part associated with pilot contamination can be expressed as Equation (30) at the top of next page, where $\widehat{\rho}_{l,k} = \sqrt{\frac{\widehat{\varpi}_{l,k}}{\varpi_k}}$. Then, the inter-cell interference caused by scattering component can be further approximated as

$$\begin{aligned}&\sum_{l=1}^{L}\frac{\widehat{\rho}_{l,k}^{2}}{\varsigma_{l,k}+1}\widehat{\boldsymbol{\omega}}_{l,k}^{H}\mathbb{E}_{\mathbf{U}_{\mathrm{S},l,k}}\left(\mathbf{U}_{\mathrm{S},l,k}^{T}\mathbf{F}_{\mathrm{RF}}\mathbf{F}_{\mathrm{RF}}^{H}\mathbf{U}_{\mathrm{S},l,k}^{*}\right)\widehat{\boldsymbol{\omega}}_{l,k} \\ \approx&\sum_{l=1}^{L}\frac{\widehat{\rho}_{l,k}^{2}}{\varsigma_{l,k}+1}\widehat{\boldsymbol{\omega}}_{l,k}^{H}\mathrm{tr}\left(\mathbf{F}_{\mathrm{RF}}\mathbf{F}_{\mathrm{RF}}^{H}\right)\mathbf{I}_{\mathrm{P}}\widehat{\boldsymbol{\omega}}_{l,k} = \sum_{l=1}^{L}\frac{\widehat{\rho}_{l,k}^{2}}{\varsigma_{l,k}+1}N. \qquad (31)\end{aligned}$$

$$\begin{aligned}
&\frac{1}{\varpi_k}\mathbb{E}_{\mathbf{U}_{l,k}}\left[\sum_{l=1}^{L}\left(\widehat{\boldsymbol{\omega}}_{l,k}^H\mathbf{U}_{l,k}^T\mathbf{F}_{\mathrm{RF}}\mathbf{F}_{\mathrm{RF}}^H\mathbf{U}_{l,k}^*\widehat{\boldsymbol{\omega}}_{l,k}\right)\right]\\
=&\mathbb{E}_{\mathbf{U}_{l,k}}\left[\sum_{l=1}^{L}\widehat{\boldsymbol{\omega}}_{l,k}^H\left(\sqrt{\frac{\varsigma_{l,k}\widehat{\rho}_{l,k}^2}{\varsigma_{l,k}+1}}\mathbf{U}_{\mathrm{SAoA},l,k}^T+\sqrt{\frac{\widehat{\rho}_{l,k}^2}{\varsigma_{l,k}+1}}\mathbf{U}_{\mathrm{S},l,k}^T\right)\mathbf{F}_{\mathrm{RF}}\mathbf{F}_{\mathrm{RF}}^H\left(\sqrt{\frac{\varsigma_{l,k}\widehat{\rho}_{l,k}^2}{\varsigma_{l,k}+1}}\mathbf{U}_{\mathrm{SAoA},l,k}^*+\sqrt{\frac{\widehat{\rho}_{l,k}^2}{\varsigma_{l,k}+1}}\mathbf{U}_{\mathrm{S},l,k}^*\right)\widehat{\boldsymbol{\omega}}_{l,k}\right]\\
\approx&\underbrace{\mathbb{E}_{\mathbf{U}_{\mathrm{SAoA},l,k}}\left[\sum_{l=1}^{L}\frac{\widehat{\rho}_{l,k}^2\varsigma_{l,k}}{\varsigma_{l,k}+1}\widehat{\boldsymbol{\omega}}_{l,k}^H\mathbf{U}_{\mathrm{SAoA},l,k}^T\mathbf{F}_{\mathrm{RF}}\mathbf{F}_{\mathrm{RF}}^H\mathbf{U}_{\mathrm{SAoA},l,k}^*\widehat{\boldsymbol{\omega}}_{l,k}\right]}_{\text{Inter}-\text{cell interference caused by strongest AoA components}}+\underbrace{\mathbb{E}_{\mathbf{U}_{\mathrm{S},l,k}}\left[\sum_{l=1}^{L}\frac{\widehat{\rho}_{l,k}^2}{\varsigma_{l,k}+1}\widehat{\boldsymbol{\omega}}_{l,k}^H\mathbf{U}_{\mathrm{S},l,k}^T\mathbf{F}_{\mathrm{RF}}\mathbf{F}_{\mathrm{RF}}^H\mathbf{U}_{\mathrm{S},l,k}^*\widehat{\boldsymbol{\omega}}_{l,k}\right]}_{\text{Inter}-\text{cell interference caused by scattering components}}.
\end{aligned}\tag{30}$$

$$\begin{aligned}
&\mathbb{E}_{\mathbf{U}_{\mathrm{SAoA},l,k}}\left[\sum_{l=1}^{L}\left(\frac{\widehat{\rho}_{l,k}^2\varsigma_{l,k}}{\varsigma_{l,k}+1}\widehat{\boldsymbol{\omega}}_{l,k}^H\mathbf{U}_{\mathrm{SAoA},l,k}^T\mathbf{F}_{\mathrm{RF}}\mathbf{F}_{\mathrm{RF}}^H\mathbf{U}_{\mathrm{SAoA},l,k}^*\widehat{\boldsymbol{\omega}}_{l,k}\right)\right]\\
=&\mathbb{E}_{\mathbf{U}_{\mathrm{SAoA},l,k}}\left[\sum_{l=1}^{L}\left[\frac{\widehat{\rho}_{l,k}^2\varsigma_{l,k}}{\varsigma_{l,k}+1}\left(\mathbf{u}_{\mathrm{SAoA},l,k}^{\mathrm{BS}}\right)^T\mathbf{F}_{\mathrm{RF}}\mathbf{F}_{\mathrm{RF}}^H\left(\mathbf{u}_{\mathrm{SAoA},l,k}^{\mathrm{BS}}\right)^*\mathbf{u}_{\mathrm{SAoA},l,k}^T\widehat{\boldsymbol{\omega}}_{l,k}\widehat{\boldsymbol{\omega}}_{l,k}^H\mathbf{u}_{\mathrm{SAoA},l,k}^*\right]\right].
\end{aligned}\tag{32}$$

$$\begin{aligned}
&\mathbb{E}_{\mathbf{U}_{\mathrm{SAoA},l,k}}\left[\sum_{l=1}^{L}\left[\frac{\widehat{\rho}_{l,k}^2\varsigma_{l,k}}{\varsigma_{l,k}+1}\left(\mathbf{u}_{\mathrm{SAoA},l,k}^{\mathrm{BS}}\right)^T\mathbf{F}_{\mathrm{RF}}\mathbf{F}_{\mathrm{RF}}^H\left(\mathbf{u}_{\mathrm{SAoA},l,k}^{\mathrm{BS}}\right)^*\mathbf{u}_{\mathrm{SAoA},l,k}^T\widehat{\boldsymbol{\omega}}_{l,k}\widehat{\boldsymbol{\omega}}_{l,k}^H\mathbf{u}_{\mathrm{SAoA},l,k}^*\right]\right]\\
=&\mathbb{E}_{\phi_{l,k},\phi_{l,k},\theta_{l,k},\theta_i}\left\{\sum_{l=1}^{L}\left[\frac{\widehat{\rho}_{l,k}^2\varsigma_{l,k}}{\varsigma_{l,k}+1}G_{\mathrm{act},P}[\cos(\phi_{l,k})-\cos(\Delta\phi_{l,k})]\sum_{i=1}^{N}G_{\mathrm{act},M}[\cos(\theta_{l,k})-\cos(\theta_i)]\right]\right\}.
\end{aligned}\tag{39}$$

Now, we would like to approximate the inter-cell interference caused by the multi-cell strongest AoA components in Equation (32) at the top of this page. In Equation (32), we have

$$\widehat{\boldsymbol{\omega}}_{l,k}^H=\frac{1}{\sqrt{P}}\left[\ 1,\ldots,e^{j2\pi(P-1)\frac{d}{\lambda}\cos(\phi_{l,k})}\ \right]^*\ \text{and}\tag{33}$$

$$\mathbf{u}_{\mathrm{SAoA},l,k}^*=\left[\ 1,\ldots,\ e^{-j2\pi(P-1)\frac{d}{\lambda}\cos(\Delta\phi_{l,k})}\ \right]^H,\tag{34}$$

where variables $\phi_{l,k}\in[0,\ \pi]$ is the angle of incidence of the strongest AoA path at antenna arrays of user $k$ in cell $l$, and $\Delta\phi_{l,k}\in[0,\ \pi]$ is the angle of incidence of the inter-cell strongest AoA path at antenna arrays from user $k$ of cell $l$ to the desired BS. By defining the array gain function $G_{\mathrm{act},P}[x]$, cf. [61], as

$$G_{\mathrm{act},P}[x]=\frac{\left\{\sin\left[P\pi\frac{d}{\lambda}(x)\right]\right\}^2}{P\left\{\sin\left[\pi\frac{d}{\lambda}(x)\right]\right\}^2},\tag{35}$$

where $\frac{d}{\lambda}=\frac{1}{2}$. Then, we have:

$$\begin{aligned}
&\mathbf{u}_{\mathrm{SAoA},l,k}^T\widehat{\boldsymbol{\omega}}_{l,k}\widehat{\boldsymbol{\omega}}_{l,k}^H\mathbf{u}_{\mathrm{SAoA},l,k}^*\\
=&G_{\mathrm{act},P}[\cos(\phi_{l,k})-\cos(\Delta\phi_{l,k})]\\
=&\frac{\left\{\sin\left[P\frac{\pi}{2}\left(\cos(\phi_{l,k})-\cos(\Delta\phi_{l,k})\right)\right]\right\}^2}{P\left\{\sin\left[\frac{\pi}{2}\left(\cos(\phi_{l,k})-\cos(\Delta\phi_{l,k})\right)\right]\right\}^2}\\
=&\frac{P\left\{\mathrm{sinc}\left[P\frac{\pi}{2}\left(\cos(\phi_{l,k})-\cos(\Delta\phi_{l,k})\right)\right]\right\}^2}{\left\{\mathrm{sinc}\left[\frac{\pi}{2}\left(\cos(\phi_{l,k})-\cos(\Delta\phi_{l,k})\right)\right]\right\}^2}.
\end{aligned}\tag{36}$$

Similarly, we can have following preliminaries, i.e.,

$$\left(\mathbf{u}_{\mathrm{SAoA},l,k}^{\mathrm{BS}}\right)^T=\left[\ 1,\ldots,\ e^{-j2\pi(M-1)\frac{d}{\lambda}\cos(\theta_{l,k})}\ \right]\ \text{and}$$
$$\widehat{\boldsymbol{\nu}}_i=\frac{1}{\sqrt{M}}\left[\ 1,\ldots,\ e^{j2\pi(M-1)\frac{d}{\lambda}\cos(\theta_i)}\ \right]^T.\tag{37}$$

Based on these aforementioned expressions, we rewrite $\left(\mathbf{u}_{\mathrm{SAoA},l,k}^{\mathrm{BS}}\right)^T\mathbf{F}_{\mathrm{RF}}\mathbf{F}_{\mathrm{RF}}^H\left(\mathbf{u}_{\mathrm{SAoA},l,k}^{\mathrm{BS}}\right)^*$ as

$$\begin{aligned}
&\left[\ \left(\mathbf{u}_{\mathrm{SAoA},l,k}^{\mathrm{BS}}\right)^T\widehat{\boldsymbol{\nu}}_1,\ldots,\left(\mathbf{u}_{\mathrm{SAoA},l,k}^{\mathrm{BS}}\right)^T\widehat{\boldsymbol{\nu}}_N\ \right]\\
&\qquad\times\left[\ \widehat{\boldsymbol{\nu}}_1^H\left(\mathbf{u}_{\mathrm{SAoA},l,k}^{\mathrm{BS}}\right)^*,\ldots,\widehat{\boldsymbol{\nu}}_N^H\left(\mathbf{u}_{\mathrm{SAoA},l,k}^{\mathrm{BS}}\right)^*\ \right]^T\\
=&\sum_{i=1}^{N}\frac{\left\{\sin\left[\frac{\pi}{2}M\left(\cos(\theta_{l,k})-\cos(\theta_i)\right)\right]\right\}^2}{M\left\{\sin\left[\frac{\pi}{2}\cos(\theta_{l,k})-\cos(\theta_i)\right]\right\}^2}\\
=&\sum_{i=1}^{N}G_{\mathrm{act},M}[\cos(\theta_{l,k})-\cos(\theta_i)].
\end{aligned}\tag{38}$$

Then, Equation (32) can be re-expressed as Equation (39) at the top of this page, where $\cos(\phi_{l,k})$, $\cos(\Delta\phi_{l,k})$, $\cos(\theta_{l,k})$, and $\cos(\theta_i)$ are uniformly distributed over $[-1,\ 1]$. Besides, they are independent with each other. Exploiting the periodic property of function $e^{j2\pi x}$, the linear antenna array gain $G_{\mathrm{act},P}[\cos(\phi_{l,k})-\cos(\Delta\phi_{l,k})]$ is equal in distribution to $G_{\mathrm{act},P}[\mu_{l,k}]$ and $G_{\mathrm{act},M}[\cos(\theta_{l,k})-\cos(\theta_i)]$ is equal in distribution to $G_{\mathrm{act},M}[\epsilon_{l,k,i}]$, where $\mu_{l,k}$, $k\in\{1,\ldots,N\}$, and $\epsilon_{l,k,i}$, $i\in\{1,\ldots,N\}$ are uniformly distributed over $[-1,\ 1]$ (Lemma 1 of [65]). In addition, $\mu_{l,k}$ and $\epsilon_{l,k,i}$ is independent

$$
\begin{aligned}
&\sum_{l=1}^{L}\frac{\widehat{\rho}_{l,k}^{2}\varsigma_{l,k}}{\varsigma_{l,k}+1}\mathbb{E}_{\phi_{l,k},\phi_{l,k}}\left[G_{\mathrm{act},P}[\cos\left(\phi_{l,k}\right)-\cos\left(\Delta\phi_{l,k}\right)]\right]\sum_{i=1}^{N}\mathbb{E}_{\theta_{l,k},\theta_i}\left[G_{\mathrm{act},M}[\cos\left(\theta_{l,k}\right)-\cos\left(\theta_i\right)]\right]\\
=&\sum_{l=1}^{L}\frac{\widehat{\rho}_{l,k}^{2}\varsigma_{l,k}}{\varsigma_{l,k}+1}\mathbb{E}_{\mu_{l,k}}\left[G_{\mathrm{act},P}[\mu_{l,k}]\right]\mathbb{E}_{\epsilon_{l,k,i}}\left(G_{\mathrm{act},M}[\epsilon_{l,k,i}]\right)\\
=&\sum_{l=1}^{L}\frac{\widehat{\rho}_{l,k}^{2}\varsigma_{l,k}}{\varsigma_{l,k}+1}\mathbb{E}_{\mu_{l,k}}\left(\frac{\left[\mathrm{sinc}\left(\frac{\pi}{2}P\mu_{l,k}\right)\right]^{2}P}{\left[\mathrm{sinc}\left(\frac{\pi}{2}\mu_{l,k}\right)\right]^{2}}\right)\mathbb{E}_{\epsilon_{l,k,i}}\left(\sum_{i=1}^{N}\frac{\left[\mathrm{sinc}\left(\frac{\pi}{2}M\epsilon_{l,k,i}\right)\right]^{2}M}{\left[\mathrm{sinc}\left(\frac{\pi}{2}\epsilon_{l,k,i}\right)\right]^{2}}\right)\\
\overset{(b)}{\geqslant}&\sum_{l=1}^{L}\frac{\widehat{\rho}_{l,k}^{2}\varsigma_{l,k}}{\varsigma_{l,k}+1}\mathbb{E}_{\mu_{l,k}}\left(\left[\mathrm{sinc}\left(\frac{\pi}{2}P\mu_{l,k}\right)\right]^{2}P\right)\mathbb{E}_{\epsilon_{l,k,i}}\left(\sum_{i=1}^{N}\left[\mathrm{sinc}\left(\frac{\pi}{2}M\epsilon_{l,k,i}\right)\right]^{2}M\right)\overset{(c)}{\approx}\sum_{l=1}^{L}\frac{\widehat{\rho}_{l,k}^{2}\varsigma_{l,k}}{\varsigma_{l,k}+1}N. \qquad (40)
\end{aligned}
$$

$$
\begin{aligned}
\Omega_k \geqslant & \underbrace{E_s\mathbb{E}_{\mathbf{H}_{\mathrm{SAoA},l,k}}\left[\sum_{l=1}^{L}\widetilde{\varpi}_{l,k}\left(\widehat{\beta}_l\right)^2\left(\frac{\varsigma_{l,k}}{\varsigma_{l,k}+1}\right)\widehat{\boldsymbol{\omega}}_k^H\mathbf{H}_{\mathrm{SAoA},l,k}^T\mathbf{F}_{\mathrm{RF},l}\widehat{\mathbf{W}}_{\mathrm{eq},l}\widehat{\mathbf{W}}_{\mathrm{eq},l}^H\mathbf{F}_{\mathrm{RF},l}^H\mathbf{H}_{\mathrm{SAoA},l,k}^*\widehat{\boldsymbol{\omega}}_k\right]}_{\text{Inter}-\text{cell interfernce caused by strongest AoAs}}\\
&+\underbrace{E_s\mathbb{E}_{\mathbf{H}_{\mathrm{S},l,k}}\left[\sum_{l=1}^{L}\widetilde{\varpi}_{l,k}\left(\widehat{\beta}_l\right)^2\left(\frac{1}{\varsigma_{l,k}+1}\right)\widehat{\boldsymbol{\omega}}_k^H\mathbf{H}_{\mathrm{S},l,k}^T\mathbf{F}_{\mathrm{RF},l}\widehat{\mathbf{W}}_{\mathrm{eq},l}\widehat{\mathbf{W}}_{\mathrm{eq},l}^H\mathbf{F}_{\mathrm{RF},l}^H\mathbf{H}_{\mathrm{S},l,k}^*\widehat{\boldsymbol{\omega}}_k\right]}_{\text{Inter}-\text{cell interference caused by scattering component}}. \qquad (43)
\end{aligned}
$$

with each other. Thus, we re-express Equation (39) as Equation (40) at the top of this page. In (b), we exploit the fact that

$$[\mathrm{sinc}(x)]^2=\left(\frac{\sin x}{x}\right)^2\leqslant 1. \qquad (41)$$

In (c), we explore the law of integration of sinc function for the number of antennas $M$ is sufficiently large, i.e.,

$$
\begin{aligned}
&\mathbb{E}_{\epsilon_{l,k,i}}\left[\left[\mathrm{sinc}\left(\frac{\pi}{2}M\epsilon_{l,k,i}\right)\right]^2M\right]\\
=&M\int_{-1}^{1}\frac{1}{2}\left[\mathrm{sinc}\left(\frac{\pi}{2}M\epsilon_{l,k,i}\right)\right]^2d\epsilon_{l,k,i}\\
=&\frac{M}{M\pi}\int_{-1}^{1}\left[\mathrm{sinc}\left(\frac{\pi}{2}M\epsilon_{l,k,i}\right)\right]^2d\frac{\pi}{2}M\epsilon_{l,k,i}\\
\overset{M\to\infty}{\approx}&\frac{1}{\pi}\int_{-\infty}^{\infty}[\mathrm{sinc}\left(\chi\right)]^2d\chi=1, \qquad (42)
\end{aligned}
$$

where $\chi=\frac{\pi}{2}M\epsilon_{l,k,i}$. We substitute Equation (31) and (40) into (29), the expression in (15) comes immediately after some straightforward mathematical manipulation.

### B. Proof of Lemma 2

The inter-cell downlink transmission interference is given by Equation (44) at the top of next page. Following similar approaches shown in Equations (32)−(38), we can re-express the inter-cell interference caused by the strongest AoA components as where $i\in\{1,\ldots,N\}$, $\widehat{\eta}_{l,k}$ and $\tau_{l,k,i}$ are also independent uniformly distributed in $[-1,\,1]$. In addition, variables $\Delta\kappa_{l,k}\in[0,\,\pi]$ is the angle of incidence of the inter-cell strongest AoA path at antenna arrays of user $k$ in the desired cell from the BS in cell $l$, and $\psi_{l,k}\in[0,\,\pi]$ is the angle of incidence of the inter-cell strongest AoA path at antenna arrays of the BS of cell $l$ for user $k$. Further, in the large number of antennas regime, $\mathbb{E}_{\tau_{l,k,i}}\left[\mathbf{F}_{\mathrm{RF},l}^H\left(\mathbf{h}_{\mathrm{SAoA},l,k}^{\mathrm{BS}}\right)^*\left(\mathbf{h}_{\mathrm{SAoA},l,k}^{\mathrm{BS}}\right)^T\mathbf{F}_{\mathrm{RF},l}\right]$, $i\in\{1,\ldots,N\}$, can be approximated as Equation (45) at the top of next page. In the large number of antennas regime, $\mathbb{E}_{\tau_{l,k,i}}\left[\frac{\mathrm{sinc}^2\left[M\frac{\pi}{2}(\tau_{l,k,i})\right]M}{\mathrm{sinc}^2\left[\frac{\pi}{2}(\tau_{l,k,i})\right]}\right]$, $i\in\{1,\ldots,N\}$, can be asymptotically approximated as

$$\mathbb{E}_{\tau_{l,k,i}}\left[\frac{\mathrm{sinc}^2\left[M\frac{\pi}{2}\left(\tau_{l,k,i}\right)\right]M}{\mathrm{sinc}^2\left[\frac{\pi}{2}\left(\tau_{l,k,i}\right)\right]}\right]\approx 1. \qquad (46)$$

In the large number of antennas regime, $i\neq j$, $i,\,j\in\{1,\ldots,N\}$, $\mathbb{E}_{\tau_{l,k,i},\tau_{l,k,j}}\left[\frac{\sin\left[M\frac{\pi}{2}(\tau_{l,k,i})\right]}{\sqrt{M}\sin\left[\frac{\pi}{2}(\tau_{l,k,i})\right]}\frac{\sin\left[M\frac{\pi}{2}(\tau_{l,k,j})\right]}{\sqrt{M}\sin\left[\frac{\pi}{2}(\tau_{l,k,j})\right]}\right]$ can be approximated as

$$\mathbb{E}_{\tau_{l,k,i}}\left[\frac{\sin\left[M\frac{\pi}{2}\left(\tau_{l,k,i}\right)\right]}{\sqrt{M}\sin\left[\frac{\pi}{2}\left(\tau_{l,k,i}\right)\right]}\right]\mathbb{E}_{\tau_{l,k,j}}\left[\frac{\sin\left[M\frac{\pi}{2}\left(\tau_{l,k,j}\right)\right]}{\sqrt{M}\sin\left[\frac{\pi}{2}\left(\tau_{l,k,j}\right)\right]}\right]\approx 0. \qquad (47)$$

Thus, in the large number of antennas regime, $i\in\{1,\ldots,N\}$, Equation (45) can be asymptotically approximated as

$$\mathbb{E}_{\tau_{l,k,i}}\left[\mathbf{F}_{\mathrm{RF},l}^H\left(\mathbf{h}_{\mathrm{SAoA},l,k}^{\mathrm{BS}}\right)^*\left(\mathbf{h}_{\mathrm{SAoA},l,k}^{\mathrm{BS}}\right)^T\mathbf{F}_{\mathrm{RF},l}\right]\approx\mathbf{I}_{\mathrm{N}}. \qquad (48)$$

Thus, Equation (44) can be asymptotically approximated as Equation (50) at the top of next page. On the other hand, we re-express the inter-cell interference caused by scattering components in the large number of antennas regime as

$$
\begin{aligned}
&\sum_{l=1}^{L}\widetilde{\boldsymbol{\omega}}_k^H\frac{\left(\frac{\widetilde{\varpi}_{l,k}}{\varsigma_{l,k}+1}\right)}{\mathrm{tr}\left(\widehat{\mathbf{W}}_{\mathrm{eq},l}\widehat{\mathbf{W}}_{\mathrm{eq},l}^H\right)}\mathrm{tr}\left(\mathbf{F}_{\mathrm{RF},l}\widehat{\mathbf{W}}_{\mathrm{eq},l}\widehat{\mathbf{W}}_{\mathrm{eq},l}^H\mathbf{F}_{\mathrm{RF},l}^H\right)\widetilde{\boldsymbol{\omega}}_k\\
\overset{M\to\infty}{\approx}&\sum_{l=1}^{L}\left(\frac{\widetilde{\varpi}_{l,k}}{\varsigma_{l,k}+1}\right). \qquad (50)
\end{aligned}
$$

$$\mathbb{E}_{\mathbf{H}_{\mathrm{SAoA},l,k}}\left\{\sum_{l=1}^{L}\widetilde{\varpi}_{l,k}\left(\widehat{\beta}_l\right)^2\left(\frac{\varsigma_{l,k}}{\varsigma_{l,k}+1}\right)\mathrm{tr}\left[\widetilde{\boldsymbol{\omega}}_k^H\mathbf{H}_{\mathrm{SAoA},l,k}^T\mathbf{F}_{\mathrm{RF},l}\widehat{\mathbf{W}}_{\mathrm{eq},l}\widehat{\mathbf{W}}_{\mathrm{eq},l}^H\mathbf{F}_{\mathrm{RF},l}^H\mathbf{H}_{\mathrm{SAoA},l,k}^*\widetilde{\boldsymbol{\omega}}_k\right]\right\}$$
$$=\sum_{l=1}^{L}\left\{\frac{\widetilde{\varpi}_{l,k}\left(\widehat{\beta}_l\right)^2\varsigma_{l,k}}{\varsigma_{l,k}+1}\mathbb{E}_{\widehat{\eta}_{l,k}}\left[\frac{\mathrm{sinc}^2\left[P\frac{\pi}{2}(\widehat{\eta}_{l,k})\right]P}{\mathrm{sinc}^2\left[\frac{\pi}{2}(\widehat{\eta}_{l,k})\right]}\right]\mathrm{tr}\left[\widehat{\mathbf{W}}_{\mathrm{eq},l}\widehat{\mathbf{W}}_{\mathrm{eq},l}^H\mathbb{E}_{\tau_{l,k,i}}\left[\mathbf{F}_{\mathrm{RF},l}^H\left(\mathbf{h}_{\mathrm{SAoA},l,k}^{\mathrm{BS}}\right)^*\left(\mathbf{h}_{\mathrm{SAoA},l,k}^{\mathrm{BS}}\right)^T\mathbf{F}_{\mathrm{RF},l}\right]\right]^T\right\}. \quad (44)$$

$$\begin{bmatrix} \mathbb{E}_{\tau_{l,k,1}}\left[\frac{\mathrm{sinc}^2\left[M\frac{\pi}{2}(\tau_{l,k,1})\right]M}{\mathrm{sinc}^2\left[\frac{\pi}{2}(\tau_{l,k,1})\right]}\right] & \cdots & \mathbb{E}_{\tau_{l,k,1},\tau_{l,k,N}}\left[\frac{\sin\left[M\frac{\pi}{2}(\tau_{l,k,1})\right]}{\sqrt{M}\sin\left[\frac{\pi}{2}(\tau_{l,k,1})\right]}\frac{\sin\left[M\frac{\pi}{2}(\tau_{l,k,N})\right]}{\sqrt{M}\sin\left[\frac{\pi}{2}(\tau_{l,k,N})\right]}\right] \\ \vdots & \ddots & \vdots \\ \mathbb{E}_{\tau_{l,k,1},\tau_{l,k,N}}\left[\frac{\sin\left[M\frac{\pi}{2}(\tau_{l,k,1})\right]}{\sqrt{M}\sin\left[\frac{\pi}{2}(\tau_{l,k,1})\right]}\frac{\sin\left[M\frac{\pi}{2}(\tau_{l,k,N})\right]}{\sqrt{M}\sin\left[\frac{\pi}{2}(\tau_{l,k,N})\right]}\right] & \cdots & \mathbb{E}_{\tau_{l,k,N}}\left[\frac{\mathrm{sinc}^2\left[M\frac{\pi}{2}(\tau_{l,k,N})\right]M}{\mathrm{sinc}^2\left[\frac{\pi}{2}(\tau_{l,k,N})\right]}\right] \end{bmatrix}. \quad (45)$$

$$\mathbb{E}_{\mathbf{H}_{\mathrm{SAoA},l,k}}\left[\sum_{l=1}^{L}\left(\frac{\widetilde{\varpi}_{l,k}\varsigma_{l,k}}{\varsigma_{l,k}+1}\right)\left[\frac{\mathrm{sinc}^2\left[P\frac{\pi}{2}(\eta_{l,k})\right]P}{\mathrm{sinc}^2\left[\frac{\pi}{2}(\eta_{l,k})\right]}\right]\widehat{\beta}^2\mathrm{tr}\left(\widehat{\mathbf{W}}_{\mathrm{eq},l}\widehat{\mathbf{W}}_{\mathrm{eq},l}^H\right)\right]$$
$$\overset{M\to\infty}{\approx}\sum_{l=1}^{L}\left(\frac{\widetilde{\varpi}_{l,k}\varsigma_{l,k}}{\varsigma_{l,k}+1}\right)\mathbb{E}_{\eta_{l,k}}\left[\frac{\mathrm{sinc}^2\left[P\frac{\pi}{2}(\eta_{l,k})\right]P}{\mathrm{sinc}^2\left[\frac{\pi}{2}(\eta_{l,k})\right]}\right]\approx\sum_{l=1}^{L}\left(\frac{\widetilde{\varpi}_{l,k}\varsigma_{l,k}}{\varsigma_{l,k}+1}\right). \quad (49)$$

By substituting Equations (49) and (50) into Equation (43), we have

$$\Omega_k \approx E_S\left[\sum_{l=1}^{L}\left(\frac{\widetilde{\varpi}_{l,k}}{\varsigma_{l,k}+1}\right)+\sum_{l=1}^{L}\left(\frac{\widetilde{\varpi}_{l,k}\varsigma_{l,k}}{\varsigma_{l,k}+1}\right)\right]$$
$$= E_s\left[\sum_{l=1}^{L}\left(\widetilde{\varpi}_{l,k}\right)\right]. \quad (51)$$

This completes the proof.

## References

[1] L. Zhao, Z. Wei, D. W. K. Ng, J. Yuan, and M. C. Reed, "Mitigating pilot contamination in multi-cell hybrid millimeter wave systems," 2018, accepted, ICC 2018. [Online]. Available: https://arxiv.org/abs/1801.09176v1

[2] L. Zhao, D. W. K. Ng, and J. Yuan, "Multi-user precoding and channel estimation for hybrid millimeter wave systems," *IEEE J. Select. Areas Commun.*, vol. 35, no. 7, pp. 1576–1590, Jul. 2017.

[3] M. Kokshoorn, H. Chen, P. Wang, Y. Li, and B. Vucetic, "Millimeter wave MIMO channel estimation using overlapped beam patterns and rate adaptation," *IEEE Trans. Signal Process.*, vol. 65, no. 3, pp. 601–616, Feb. 2017.

[4] N. Yang, L. Wang, G. Geraci, M. Elkashlan, J. Yuan, and M. D. Renzo, "Safeguarding 5G wireless communication networks using physical layer security," *IEEE Commun. Mag.*, vol. 53, no. 4, pp. 20–27, Apr. 2015.

[5] L. Dai, Z. Wang, and Z. Yang, "Spectrally efficient time-frequency training OFDM for mobile large-scale MIMO systems," *IEEE J. Select. Areas Commun.*, vol. 31, no. 2, pp. 251–263, Feb. 2013.

[6] J. Zhu, R. Schober, and V. K. Bhargava, "Secure transmission in multicell massive MIMO systems," *IEEE Trans. Wireless Commun.*, vol. 13, no. 9, pp. 4766–4781, Sep. 2014.

[7] J. A. Zhang, X. Huang, V. Dyadyuk, and Y. J. Guo, "Massive hybrid antenna array for millimeter-wave cellular communications," *IEEE Wireless Commun.*, vol. 22, no. 1, pp. 79–87, Feb. 2015.

[8] V. W. S. Wong, R. Schober, D. W. K. Ng, and L.-C. Wang, *Key Technologies for 5G Wireless Systems*. Cambridge Univ. Press, 2017.

[9] S. Akbar, Y. Deng, A. Nallanathan, M. Elkashlan, and G. K. Karagiannidi, "Massive multiuser MIMO in heterogeneous cellular networks with full duplex small cells," *IEEE Trans. Commun.*, vol. PP, no. 99, pp. 1–1, 2017.

[10] F. Sohrabi and W. Yu, "Hybrid digital and analog beamforming design for large-scale antenna arrays," *IEEE J. Select. Topics in Signal Process.*, vol. 10, no. 3, pp. 501–513, Apr. 2016.

[11] F. Rusek, D. Persson, B. Lau, E. G. Larsson, T. L. Marzetta, O. Edfors, and F. Tufvesson, "Scaling up MIMO : Opportunities and challenges with very large arrays," *IEEE Trans. Signal Process.*, vol. 30, no. 1, pp. 40–60, Jan. 2013.

[12] J. Andrews, S. Buzzi, C. Wan, S. Hanly, A. L. A. Soong, and J. Zhang, "What will 5G be?" *IEEE J. Select. Areas Commun.*, vol. 32, no. 6, pp. 1065–1082, Jun. 2014.

[13] M. Zhao, Z. Shi, and M. C. Reed, "Iterative turbo channel estimation for OFDM system over rapid dispersive fading channel," *IEEE Trans. Wireless Commun.*, vol. 7, no. 8, pp. 3174–3184, Aug. 2008.

[14] C. Lin, G. Y. Li, and L. Wang, "Subarray-based coordinated beamforming training for mmWave and sub-THz communications," *IEEE J. Select. Areas Commun.*, vol. 35, no. 9, pp. 2115–2126, Sep. 2017.

[15] G. Wang, F. Gao, W. Chen, and C. Tellambura, "Channel estimation and training design for two-way relay networks in time-selective fading environments," *IEEE Trans. Wireless Commun.*, vol. 10, no. 8, pp. 2681–2691, Aug. 2011.

[16] E. G. Larsson, O. Edfors, F. Tufvesson, and T. L. Marzetta, "Massive MIMO for next generation wireless systems," *IEEE Commun. Mag.*, vol. 52, no. 2, pp. 186–195, Feb. 2014.

[17] T. L. Marzetta, "Noncooperative cellular wireless with unlimited numbers of base station antennas," *IEEE Trans. Wireless Commun.*, vol. 9, no. 11, pp. 3590–3600, Nov. 2010.

[18] T. S. Rappaport, G. R. MacCartney, M. K. Samimi, and S. Sun, "Wideband millimeter-wave propagation measurements and channel models for future wireless communication system design," *IEEE Trans. Commun.*, vol. 63, no. 9, pp. 3029–3056, Sep. 2015.

[19] Y. Wu, R. Schober, D. W. K. Ng, C. Xiao, and G. Caire, "Secure massive MIMO transmission with an active eavesdropper," *IEEE Trans. Inf. Theory*, vol. 62, no. 7, pp. 3880–3900, Jul. 2016.

[20] Z. Wei, D. W. K. Ng, J. Yuan, and H. M. Wang, "Optimal resource allocation for power-efficient MC-NOMA with imperfect channel state information," *IEEE Trans. Commun.*, vol. 65, no. 9, pp. 3944–3961, Sep. 2017.

[21] Z. Wei, L. Zhao, J. Guo, D. W. K. Ng, and J. Yuan, "A multi-beam NOMA framework for hybrid mmwave systems," 2018, accepted, ICC 2018. [Online]. Available: https://arxiv.org/abs/1804.08303

[22] E. Björnson, E. G. Larsson, and T. L. Marzetta, "Massive MIMO: Ten myths and one critical question," *IEEE Commun. Mag.*, vol. 54, no. 2, pp. 114–123, Feb. 2016.

[23] L. Zhao, G. Geraci, T. Yang, D. W. K. Ng, and J. Yuan, "A tone-based AoA estimation and multiuser precoding for millimeter wave massive MIMO," *IEEE Trans. Commun.*, vol. 65, no. 12, pp. 5209–5225, Dec. 2017.

[24] T. E. Bogale and L. B. Le, "Massive MIMO and millimeter wave for 5G wireless HetNet: Potentials and challenges," *IEEE Veh. Technol. Mag.*, vol. 11, no. 1, pp. 64–75, Mar. 2016.

[25] K. Sakaguchi and et al, "Where, when, and how mmWave is used in 5G and beyon," 2017. [Online]. Available: https://arxiv.org/abs/1704.08131

[26] K. Sahota, "5G mmWave radio design for mobile," 2017. [Online]. Available: https://5gsummit.org/hawaii/docs/slides

[27] G. Miao and G. Song, *Energy and spectrum efficient wireless network design*. Cambridge Univ. Press, 2014.

[28] Y. Niu, Y. Li, D. Jin, L. Su, and A. V. Vasilakos, "A survey of millimeter wave (mmWave) communications for 5G: Opportunities and challenges," 2015. [Online]. Available: https://arxiv.org/abs/1502.07228

[29] V. Petrov, M. Komarov, D. Moltchanov, J. M. Jornet, and Y. K-oucheryavy, "Interference and SINR in millimeter wave and terahertz communication systems with blocking and directional antennas," *IEEE Trans. Wireless Commun.*, vol. 16, no. 3, pp. 1791–1808, Mar. 2017.

[30] A. Alkhateeb, G. Leus, and R. W. Heath, "Limited feedback hybrid precoding for multi-user millimeter wave systems," *IEEE Trans. Wireless Commun.*, vol. 14, no. 11, pp. 6481–6494, Nov. 2015.

[31] H. Xie, F. Gao, S. Zhang, and S. Jin, "A unified transmission strategy for TDD/FDD massive MIMO systems with spatial basis expansion model," *IEEE Trans. Veh. Technol.*, vol. 66, no. 4, pp. 3170–3184, Apr. 2017.

[32] J. C. Shen, J. Zhang, and K. B. Letaief, "Downlink user capacity of massive MIMO under pilot contamination," *IEEE Trans. Wireless Commun.*, vol. 14, no. 6, pp. 3183–3193, Jun. 2015.

[33] S. Wagner, R. Couillet, M. Debbah, and D. T. M. Slock, "Large system analysis of linear precoding in correlated MISO broadcast channels under limited feedback," *IEEE Trans. Inf. Theory*, vol. 58, no. 7, pp. 4509–4537, Jul. 2012.

[34] X. Gao, L. Dai, S. Han, C. L. I, and R. W. Heath, "Energy-efficient hybrid analog and digital precoding for mmWave MIMO systems with large antenna arrays," *IEEE J. Select. Areas Commun.*, vol. 34, no. 4, pp. 998–1009, Apr. 2016.

[35] M. R. Akdeniz, Y. Liu, M. K. Samimi, S. Sun, S. Rangan, T. S. Rappaport, and E. Erkip, "Millimeter wave channel modeling and cellular capacity evaluation," *IEEE J. Select. Areas Commun.*, vol. 32, no. 6, pp. 1164–1179, Jun. 2014.

[36] S. Hur, S. Baek, B. Kim, Y. Chang, A. F. Molisch, T. S. Rappaport, K. Haneda, and J. Park, "Proposal on millimeter-wave channel modeling for 5G cellular system," *IEEE J. Select. Topics in Signal Process.*, vol. 10, no. 3, pp. 454–469, Apr. 2016.

[37] T. S. Rappaport, G. R. MacCartney, S. Sun, H. Yan, and S. Deng, "Small-scale, local area, and transitional millimeter wave propagation for 5G communications," *IEEE Trans. on Antennas and Propagation*, vol. 65, no. 12, pp. 6474–6490, Dec. 2017.

[38] S. Hur, Y. J. Cho, J. Lee, N.-G. Kang, J. Park, and H. Benn, "Synchronous channel sounder using horn antenna and indoor measurements on 28 GHz," in *2014 IEEE Intern. Black Sea Conf. on Commun. and Networking (BlackSeaCom)*, May 2014, pp. 83–87.

[39] J. P. Gonzalez-Coma, J. Rodriguez-Fernandez, N. Gonzalez-Prelcic, L. Castedo, and R. W. Heath, "Channel estimation and hybrid precoding for frequency selective multiuser mmwave MIMO systems," *IEEE J. Select. Topics in Signal Process.*, pp. 1–1, 2018.

[40] V. Raghavan, S. Subramanian, J. Cezanne, A. Sampath, O. H. Koymen, and J. Li, "Single-user versus multi-user precoding for millimeter wave MIMO systems," *IEEE J. Select. Areas Commun.*, vol. 35, no. 6, pp. 1387–1401, Jun. 2017.

[41] V. Raghavan, J. Cezanne, S. Subramanian, A. Sampath, and O. Koymen, "Beamforming tradeoffs for initial UE discovery in millimeter-wave MIMO systems," *IEEE J. Select. Topics in Signal Process.*, vol. 10, no. 3, pp. 543–559, Apr. 2016.

[42] R. Magueta, D. Castanheira, A. Silva, R. Dinis, and A. Gameiro, "Hybrid iterative space-time equalization for multi-user mmw massive MIMO systems," *IEEE Trans. Commun.*, vol. 65, no. 2, pp. 608–620, Feb. 2017.

[43] J. Flordelis, F. Rusek, F. Tufvesson, E. G. Larsson, and O. Edfors, "Massive MIMO performance - TDD versus FDD: What do measurements say?" *IEEE Trans. Wireless Commun.*, vol. PP, no. 99, pp. 1–1, 2018.

[44] V. Raghavan, A. Partyka, L. Akhoondzadeh-Asl, M. A. Tassoudji, O. H. Koymen, and J. Sanelli, "Millimeter wave channel measurements and implications for PHY layer design," *IEEE Trans. Antennas and Propagation*, vol. 65, no. 12, pp. 6521–6533, Dec. 2017.

[45] J. Ko, Y. J. Cho, S. Hur, T. Kim, J. Park, A. F. Molisch, K. Haneda, M. Peter, D. J. Park, and D. H. Cho, "Millimeter-wave channel measurements and analysis for statistical spatial channel model in in-building and urban environments at 28 GHz," *IEEE Trans. Wireless Commun.*, vol. 16, no. 9, pp. 5853–5868, Sep. 2017.

[46] J. Jose, A. Ashikhmin, T. L. Marzetta, and S. Vishwanath, "Pilot contamination and precoding in multi-cell TDD systems," *IEEE Trans. Wireless Commun.*, vol. 10, no. 8, pp. 2640–2651, Aug. 2011.

[47] Y. Wu, R. Schober, D. W. K. Ng, C. Xiao, and G. Caire, "Secure massive MIMO transmission with an active eavesdropper," *IEEE Trans. Inf. Theory*, vol. 62, no. 7, pp. 3880–3900, Jul. 2016.

[48] L. Sanguinetti, A. Kammoun, and M. Debbah, "Asymptotic analysis of multicell massive MIMO over Rician fading channels," in *Proc. IEEE Intern. Conf. on Acoustics, Speech and Signal Process. (ICASSP)*, Mar. 2017, pp. 3539–3543.

[49] N. Akbar, N. Yang, P. Sadeghi, and R. A. Kennedy, "Multi-cell multiuser massive MIMO networks: User capacity analysis and pilot design," *IEEE Trans. Commun.*, vol. 64, no. 12, pp. 5064–5077, Dec. 2016.

[50] W. A. W. M. Mahyiddin, P. A. Martin, and P. J. Smith, "Performance of synchronized and unsynchronized pilots in finite massive MIMO systems," *IEEE Trans. Wireless Commun.*, vol. 14, no. 12, pp. 6763–6776, Dec. 2015.

[51] T. E. Bogale, L. B. Le, X. Wang, and L. Vandendorpe, "Pilot contamination mitigation for wideband massive MMO: Number of cells vs multipath," in *Proc. IEEE Global Telecommun. Conf. (GLOBECOM)*, Dec. 2015, pp. 1–6.

[52] H. Yin, D. Gesbert, M. Filippou, and Y. Liu, "A coordinated approach to channel estimation in large-scale multiple-antenna systems," *IEEE J. Select. Areas Commun.*, vol. 31, no. 2, pp. 264–273, Feb. 2013.

[53] H. Yin, L. Cottatellucci, D. Gesbert, R. R. Mller, and G. He, "Robust pilot decontamination based on joint angle and power domain discrimination," *IEEE Trans. Signal Process.*, vol. 64, no. 11, pp. 2990–3003, Jun. 2016.

[54] E. Björnson, J. Hoydis, and L. Sanguinetti, "Massive MIMO has unlimited capacity," *IEEE Trans. Wireless Commun.*, vol. PP, no. 99, pp. 1–1, 2017.

[55] J. G. Andrews, H. Claussen, M. Dohler, S. Rangan, and M. C. Reed, "Femtocells: Past, present, and future," *IEEE J. Select. Areas Commun.*, vol. 30, no. 3, pp. 497–508, Apr. 2012.

[56] X. Zhang, A. F. Molisch, and S.-Y. Kung, "Variable-phase-shift-based RF-baseband codesign for MIMO antenna selection," vol. 53, no. 11, pp. 4091–4103, Nov. 2005.

[57] E. Zhang and C. Huang, "On achieving optimal rate of digital precoder by RF-baseband codesign for MIMO systems," in *Proc. IEEE Veh. Techn. Conf. (VTC)*, Sep. 2014, pp. 1–5.

[58] T. E. Bogale, L. B. Le, A. Haghighat, and L. Vandendorpe, "On the number of RF chains and phase shifters, and scheduling design with hybrid analog-digital beamforming," *IEEE Trans. Wireless Commun.*, vol. 15, no. 5, pp. 3311–3326, May 2016.

[59] A. A. M. Saleh and R. Valenzuela, "A statistical model for indoor multipath propagation," *IEEE J. Select. Areas Commun.*, vol. 5, no. 2, pp. 128–137, Feb. 1987.

[60] L. Dai, X. Gao, S. Han, I. Chih-Lin, and X. Wang, "Beamspace channel estimation for millimeter-wave massive MIMO systems with lens antenna array," in *2016 IEEE/CIC Intern. Conf. on Commun. in China (ICCC)*, Jul. 2016, pp. 1–6.

[61] D. Tse and P. Viswanath, *Fundamentals of wireless communication*. Cambridge University Press, 2005.

[62] Z. Al-Daher, L. P. Ivrissimtzis, and A. Hammoudeh, "Electromagnetic modeling of high-frequency links with high-resolution terrain data," *IEEE Antennas and Wireless Propagation Lett.*, vol. 11, pp. 1269–1272, Oct. 2012.

[63] Z. Muhi-Eldee, L. P. Ivrissimtzis, and M. Al-Nuaimi, "Modeling and measurements of millimeter wave propagation in urban environments," *IET Microw. Antennas Propagation*, vol. 4, no. 9, pp. 1300–1309, Sep. 2010.

[64] A. Li and C. Masouros, "Hybrid analog-digital millimeter-wave MU-MIMO transmission with virtual path selection," *IEEE Commun. Lett.*, vol. 21, no. 2, pp. 438–441, Feb. 2017.

[65] X. Yu, J. Zhang, M. Haenggi, and K. B. Letaief, "Coverage analysis for millimeter wave networks: The impact of directional antenna arrays," *IEEE J. Select. Areas Commun.*, vol. 35, no. 7, pp. 1498–1512, Jul. 2017.